\documentclass[a4paper,11pt]{article}
\pdfoutput=1 
\usepackage{jheppub} 
\usepackage[T1]{fontenc}
\usepackage{slashed,xcolor,braket,hyperref,amsmath,amssymb,epsfig,graphicx,comment}
\usepackage[utf8]{inputenc}
\allowdisplaybreaks

\definecolor{nicered}{rgb}{0.7,0.1,0.1}
\definecolor{nicegreen}{rgb}{0.1,0.5,0.1}
\definecolor{niceblue}{rgb}{0.0,0.1,0.7}
\hypersetup{colorlinks,citecolor=nicegreen,linkcolor=nicered,urlcolor=niceblue}
\newcommand{\overbar}[1]{\mkern 3.5mu\overline{\mkern-3.5mu#1\mkern-3.5mu}\mkern 3.5mu}

\DeclareMathOperator{\im}{Im}

\newcommand{\specialcell}[1]{\ifmeasuring@#1\else\omit$\displaystyle#1$\ignorespaces\fi}

\def \beq{\begin{equation}}
\def \eeq{\end{equation}}
\def \bea{\begin{eqnarray}}
\def \eea{\end{eqnarray}}

\title{Semi-leptonic three-body proton  decay modes \\ from light-cone sum rules}

\author[1]{Ulrich Haisch}

\author[1]{and Amando Hala}

\affiliation[1]{Max Planck Institute for Physics, F{\"o}hringer Ring 6,  80805 M{\"u}nchen, Germany}   

\emailAdd{haisch@mpp.mpg.de}
\emailAdd{ahala@mpp.mpg.de}

\abstract{Using light-cone sum rule techniques, we  estimate the form factors which parametrise the hadronic matrix elements that are relevant for semi-leptonic three-body proton  decays. The obtained form factors allow us to  determine the differential rate for the decay of a proton ($p$) into a positron ($e^+$), a~neutral pion~($\pi^0$) and a graviton~($G$), which  is  the leading proton decay channel in the effective theory of gravitons and  Standard~Model particles~(GRSMEFT). The sensitivity of existing and next-generation neutrino experiments in detecting the $p \to e^+ \pi^0 G$ signature is studied and the phenomenological implications  of our computations for constraints on the effective mass scale  that suppresses the relevant baryon-number violating~GRSMEFT operator are discussed.}

\begin{document} 

\maketitle

\section{Introduction}
\label{sec:intro}

Proton decay constitutes one of the most sensitive probes of high-scale physics beyond the Standard Model~(BSM).  Most of the existing proton decay searches have focused on two-body decay channels, since transitions such as $p \to e^+ \pi^0$ that involve  a positron ($e^+)$ and a neutral pion ($\pi^0$) are generically predicted by theories of grand unification (GUTs). In a broader context, studying baryon-number violation can be interesting in the light of supersymmetric theories,  baryogenesis  and theories of quantum gravity, where the global symmetries of the Standard~Model~(SM) are expected to be broken at some level --- for a review on baryon-number violation in various BSM models see~\cite{Nath:2006ut}. The null results provided by these experiments (cf.~\cite{Takhistov:2016eqm,Heeck:2019kgr,Girmohanta:2019xya} for comprehensive summaries of the available experimental results)  together with the observation that many higher-dimensional operators violating baryon number $(B)$ by one unit induce multi-body proton decay modes, make proton decay processes with more complicated final states interesting search targets for existing and next-generation neutrino experiments like Super-Kamiokande~(SK)~\cite{Super-Kamiokande:2002weg}, Hyper-Kamiokande~(HK)~\cite{Abe:2011ts}, JUNO~\cite{JUNO:2015zny} and~DUNE~\cite{DUNE:2015lol}. 

While for  all two-body proton decays into anti-leptons and pseudoscalar mesons, lattice~QCD~(LQCD) techniques nowadays allow direct computations of the  relevant hadronic matrix elements within uncertainties of $(10-15)\%$~\cite{Gavela:1988cp,Aoki:1999tw,Tsutsui:2004qc,Aoki:2006ib,Braun:2008ur,Aoki:2013yxa,Aoki:2017puj,Yoo:2018fyn},  LQCD calculations of three-body decay  modes do not exist at present although the formalism and methodologies are in principle known~\cite{Cirigliano:2019jig}. Model estimates of three-body final-state proton decay rates are therefore available only for selected modes~\cite{Wise:1980ch} or rely on naive dimensional analysis and phase-space arguments~\cite{Heeck:2019kgr,Girmohanta:2019xya}.  In our recent article~\cite{Haisch:2021hvj} we have shown that by employing light-cone~sum~rules~(LCSRs)~\cite{Balitsky:1986st,Balitsky:1997wi,Balitsky:1988qn,Braun:1988qv,Balitsky:1989ry,Chernyak:1990ag} it is possible to reproduce the LQCD results for the hadronic matrix elements relevant for GUT-like proton decay. The goal of this work  is to apply the~LCSRs formalism developed in the latter publication to the case of semi-leptonic three-body proton decay processes. In particular, we will describe in detail the calculation of all form factors needed to compute the differential decay rate for the process~$p \to e^+ \pi^0 G$~with~$G$~denoting a graviton. This decay mode is the leading proton decay channel in the effective theory that describes the interactions of gravitons and SM particles aka GRSMEFT~\cite{Ruhdorfer:2019qmk,Durieux:2019siw}. In~a~companion paper~\cite{inprep} we will analyse a broad range of possible laboratory probes of the~GRSMEFT, showing that proton decay measurements set the nominally strongest bound on the effective mass scale  that suppresses the~GRSMEFT interactions. While in this work the focus  lies on obtaining predictions for  $p \to e^+ \pi^0 G$, the provided analytic expressions and numerical results can be used to obtain the differential decay rates of other proton decay modes with a single pion in the final state. 

This article is structured as follows. In Section~\ref{sec:setup} we introduce the relevant baryon-number violating GRSMEFT interactions and provide a suitable representation for the decay amplitude of $p \to e^+ \pi^0 G$ in terms of a leptonic and a hadronic part. In Section~\ref{sec:had} we outline the calculation of the relevant hadronic matrix elements using~LCSR techniques, while the structure of the resulting~LCSRs is discussed in~Section~\ref{sec:anatomy}.  We turn to the numerical evaluation of the~LCSRs in  Section~\ref{sec:numerics}, providing  predictions and uncertainty estimates for the proton-to-pion form factors in the physical region. The differential decay rate of $p \to e^+ \pi^0 G$ is computed in Section~\ref{sec:pheno}. This section also contains a discussion of the sensitivity of existing and next-generation neutrino experiments to the  studied  proton decay signature. Our conclusions and an outlook are presented   in~Section~\ref{sec:conclusions}. Supplementary material is relegated to a number of  appendices.
 
 \section{Preliminaries}
 \label{sec:setup}
 
Considering the case of one generation of fermions, baryon-number violation~($\slashed{B}$) is induced by only a single dimension-eight operator in the GRSMEFT~\cite{Ruhdorfer:2019qmk,Durieux:2019siw}. We write this operator in the following way 
\begin{equation}
\mathcal{L}_{\slashed{B}}^{(8)} =  c_{\slashed{B}}  \hspace{0.25mm} \epsilon^{abc} \left(d_a^T C \sigma^{\mu\nu} P_R u_b \right) \left(e^T C \sigma^{\rho\sigma} P_R u_c\right) C_{\mu\nu\rho\sigma} + \text{h.c.} \,,
\label{eq:dim8}
\end{equation}
where $u$ and $d$ denote the up- and down-quark field, $e$ is the electron field, $P_R$ projects on right-handed fields,  $\epsilon^{abc}$ is the fully antisymmetric Levi-Civita tensor with $a,b,c$ denoting colour indices,  $C$ is the charge conjugation matrix, $T$ denotes the transpose with respect to Dirac indices and $\sigma^{\mu\nu} \equiv i \left (\gamma^\mu \gamma^\nu - \gamma^\nu \gamma^\mu\right)/2$ with $\gamma_\mu$ the usual Dirac matrices. Furthermore,~$C_{\mu\nu\rho\sigma}$ represents the Weyl tensor which is the traceless part of the Riemann tensor~$R_{\mu\nu\rho\sigma}$. It takes the form
\beq \label{eq:weyl}
C_{\mu\nu\rho\sigma} = R_{\mu\nu\rho\sigma} - \left (g_{\mu [ \rho} R_{\sigma ] \nu} - g_{\nu [ \rho} R_{\sigma ] \mu} \right ) + \frac{1}{3} \hspace{0.25mm} g_{\mu [ \rho}  g_{\sigma ] \nu}  R \,,
\eeq
with $g_{\mu \nu}$ the metric tensor, $R_{\mu \nu}$ the Ricci tensor, $R$ the Ricci scalar and the brackets denote index antisymmetrisation,~i.e.~$X_{[\mu} Y_{\nu]} = (X_\mu Y_\nu - X_\nu Y_\mu)/2$. Notice that the Wilson coefficient $c_{\slashed{B}}$ entering~(\ref{eq:dim8}) carries mass dimension $-4$. 

The amplitude for the decay $p (p_p) \to e^+ (p_e) \pi^0 (p_\pi) G (p)$ can be written as
\begin{equation}
\mathcal{A}\left( p\rightarrow e^+ \pi^0  G\right) = -2 \hspace{0.125mm} \kappa  \hspace{0.25mm}  c_{\slashed{B}}  \,\varepsilon_{\mu \rho}^\ast(p,\lambda)\, p_\sigma p_\nu \, \bar{v}_e^c (p_e)\,P_R \,\sigma^{\rho\sigma}\, H^{\mu \nu}(p_p,q) \hspace{0.5mm} u_p(p_p) \,,
\label{eq:pdecampl}
\end{equation}
where  $\kappa \equiv 2/\overbar{M}_{\hspace*{-1.5pt} {P}}$ with the reduced Planck mass $\overbar{M}_{\hspace*{-1.5pt} {P}} \equiv 1/\sqrt{8 \hspace{0.125mm} \pi \hspace{0.125mm}  G_N} \simeq 2.435 \cdot 10^{18}$ GeV, and $G_N \simeq 6.709 \cdot 10^{-39} \text{ GeV}^{-2}$~\cite{Zyla:2020zbs} is the gravitational or Newton constant. In addition,~$u_p(p_p)$~denotes the spinor of the proton with four-momentum $p_p$, $\bar{v}_e^c (p_e)$ is the charge conjugate anti-spinor of the electron with momentum $p_e$ and $\varepsilon_{\mu \rho}^\ast(p,\lambda)$ denotes the conjugate of the polarisation tensor of the graviton with four-momentum $p$ and polarisation $\lambda$. The~variable~$q \equiv p_p-p_\pi= p+p_e$ denotes the four-momentum transfer from the proton to the neutral pion, and enters $\mathcal{A}\left( p\rightarrow e^+ \pi^0  G\right)$  through the hadronic tensor $H_{\mu\nu}(p_p,q)$. Notice that in order to obtain~(\ref{eq:pdecampl})   the gauge of the graviton is chosen such   that the polarisation tensor is transverse and traceless,~i.e.~the following terms can be omitted in the decomposition of the decay amplitude~(\ref{eq:pdecampl})
\begin{equation}
p^\mu \varepsilon^\ast_{\mu \rho} (p,\lambda)  = 0 \,,\qquad  \varepsilon^{\ast\, \mu}_{\mu} (p,\lambda) =0 \,.
\label{eq:transverse}
\end{equation}

\section{Hadronic form factors}
\label{sec:had}

In what follows we  discuss the necessary steps  to calculate the hadronic tensor~$H_{\mu\nu}(p_p,q)$ with the help of LCSRs in QCD. We employ the notation and the conventions introduced in our earlier article~\cite{Haisch:2021hvj}. The starting point for the sum rules is the correlation function
\begin{equation}
\Pi_{\mu\nu} (p_p,q) = i \int \! d^4x \, e^{iqx} \braket{\pi^0(p_\pi)| \, T \left[ Q_{\mu\nu}(x) \bar{\eta}(0) \right] |0} \,,
\end{equation}
where $T$ denotes time ordering and the current $\eta$ is a combination of three quark fields that interpolates the proton 
\begin{equation}
\braket{0| \eta(0) | p(p_p)} =m_p \lambda_p u_p( p_p) \,.
\label{eq:protoncoupl}
\end{equation}
Here $m_p \simeq 938 \, {\rm MeV}$ is the proton mass and $\lambda_p$ denotes the coupling strength of the current $\eta$ to the physical proton state. The~strongly-interacting part of the dimension-eight operator~(\ref{eq:dim8}) is encoded by 
\begin{equation}
Q_{\mu\nu} (x) \equiv \epsilon^{abc} \left(d_a^T(x)  C \sigma_{\mu\nu} P_R u_b(x)\right) P_R u_c(x)\,.
\end{equation}

By following the standard procedure the hadronic representation of the sum rules can be cast into the form
\begin{equation}
\Pi^{\rm had}_{\mu\nu} (p_p,q) = -\frac{m_p}{p_p^2 - m_p^2 + i \epsilon} \hspace{0.5mm} \lambda_p \, H_{\mu\nu}(p_p,q) \left(\slashed{p}_p + m_p \right) + \ldots \,,
\label{eq:corhadgeneral}
\end{equation}
with $\epsilon >0$ and infinitesimal and the ellipsis denotes contributions from heavier states,~i.e.~excited states  and the continuum.  The hadronic tensor  $H_{\mu\nu}(p_p,q)$ that characterises the $p \rightarrow \pi^0$ transition can be parameterised by four independent form factors $w_n$  with $n=1,2,3,4$ in the following way  
\begin{equation}
\begin{split}
H^{\mu\nu}(p_p,q)  \hspace{0.5mm} u_p(p_p) & \equiv \braket{\pi^0(p_\pi) |  \, \epsilon^{abc} \left(d_a^T C \sigma^{\mu\nu} P_R u_b \right) P_R u_c \,  | p(p_p)} \\[2mm]
& = P_R \bigg[ \left( i \epsilon^{\mu\nu p_p q} + 2 \hspace{0.25mm} p_p^{[\mu} q^{\nu]} \right) \frac{w_1}{m_p^2}  +
 i \sigma^{p_p q} \left( i \epsilon^{\mu\nu p_p q} + 2\hspace{0.25mm} p_p^{[\mu} q^{\nu]} \right) \frac{w_2}{m_p^4} \\[2mm]
& \qquad \quad + 2 \hspace{0.25mm} i \left( p_p^{[\mu} \sigma^{\nu] q} - q^{[\mu} \sigma^{\nu] p_p} \right) \frac{w_3}{m_p^2} +  i \sigma^{\mu\nu} w_4 \bigg] \, u_p(p_p) \,.
\end{split}
\label{eq:onshelltensor}
\end{equation}
Here the proton spinor is understood to be on-shell, $\epsilon^{\mu\nu\rho\sigma}$ is the fully antisymmetric Levi-Civita tensor with  $\epsilon^{0123} = 1$ and we have introduced the abbreviations $\sigma^{\mu p} \equiv \sigma^{\mu\nu} \hspace{0.5mm}  p_\nu$, $\sigma^{p q} \equiv \sigma^{\mu\nu} \hspace{0.5mm}  p_\mu q_\nu$ and $\epsilon^{\mu \nu p q} \equiv \epsilon^{\mu\nu\rho \sigma} \hspace{0.5mm}  p_\rho q_\sigma$.  We add that after making use of the Dirac equation and algebraic identities the result~(\ref{eq:onshelltensor}) matches the  decomposition provided in~\cite{Pire:2005ax,Pire:2021hbl}.  

The hadronic representation of the correlation function~\eqref{eq:corhadgeneral} can be written as
\bea
\begin{split}
\Pi^{\rm had}_{\mu\nu} (p_p,q) = P_R \bigg[& \frac{1}{m_p^2} \left( i \epsilon^{\mu\nu p_p q} + 2 \hspace{0.25mm} p_p^{[\mu} q^{\nu]} \right) \Pi^{\rm had}_{S} + \frac{\slashed{q}}{m_p^3} \left( i \epsilon^{\mu\nu p_p q} + 2 \hspace{0.25mm} p_p^{[\mu} q^{\nu]} \right) \Pi^{\rm had}_{A_1} \\[2mm]
& + \frac{\slashed{p}_p}{m_p^3} \left( i \epsilon^{\mu\nu p_p q} + 2 \hspace{0.25mm} p_p^{[\mu} q^{\nu]} \right) \Pi^{\rm had}_{A_2} + \frac{1}{m_p} \left( i \gamma_\rho \epsilon^{\mu\nu \rho q} +2 \hspace{0.25mm} \gamma^{[\mu} q^{\nu]} \right) \Pi^{\rm had}_{A_3} \\[2mm]
& + \frac{1}{m_p} \left( i \gamma_\rho \epsilon^{\mu\nu \rho p_p} + 2 \hspace{0.25mm} \gamma^{[\mu} p_p^{\nu]} \right) \Pi^{\rm had}_{A_4} + i \sigma^{\mu\nu} \hspace{0.125mm} \Pi^{\rm had}_{T_1} \\[2mm]
& + \frac{2 \hspace{0.25mm} i}{m_p^2} \left( p_p^{[\mu} \sigma^{\nu] q} - q^{[\mu} \sigma^{\nu] p_p} \right) \Pi^{\rm had}_{T_2} + \frac{i \sigma^{p_p q}}{m_p^4} \left( i \epsilon^{\mu\nu p_p q} + 2 \hspace{0.25mm} p_p^{[\mu} q^{\nu]} \right) \Pi^{\rm had}_{T_3} \bigg] \,, \quad 
\end{split}
\label{eq:corhadexpl}
\eea
where $\epsilon^{\mu \nu \rho p} \equiv \epsilon^{\mu\nu\rho \sigma} \hspace{0.5mm}  p_\sigma$. The eight Dirac structures in~(\ref{eq:corhadexpl}) can be used to derive LCSRs for the four form factors $w_n$ or combinations of them. The corresponding scalar functions~$\Pi^{\rm had}_{\alpha}$ with $\alpha=S,A_1,A_2,A_3,A_4,T_1,T_2,T_3$ depend only on the square $p_p^2$ of the proton four-momentum  and on the square $Q^2 \equiv -q^2$ of the four-momentum transfer between the proton and the neutral pion. They can be expressed as dispersive integrals as follows 
\begin{equation}
\Pi^{\rm had}_{\alpha} (p_p^2,Q^2) = \int_{m_p^2}^\infty \! ds \; \frac{\rho^{\rm had}_{\alpha}(s,Q^2)}{s-p_p^2} \,,
\label{eq:haddisp}
\end{equation}
where
\begin{equation}
\rho^{\rm had}_{\alpha}(s,Q^2) \equiv \frac{1}{\pi} \hspace{0.25mm} \im \Pi^{\rm had}_{\alpha} (s+i \epsilon,Q^2)
\end{equation}
are spectral densities. In this way the ground-state contribution can be separated from the contribution due to heavier states denoted by $ \rho^{\rm cont}_{\alpha}(s,Q^2)$:
\begin{equation} \label{eq:sep}
\rho^{\rm had}_{\alpha}(s,Q^2) = \lambda_p \hspace{0.25mm} m_p^2 \, \delta \!\left(s-m_p^2\right) W_{\alpha}(s,Q^2) + \rho^{\rm cont}_{\alpha}(s,Q^2) \,.
\end{equation}
On-shell,~i.e.~for  $s=m_p^2$ , the functions $W_{\alpha} (Q^2) \equiv W_{\alpha} (m_p^2, Q^2)$ take the following form 
\bea
\begin{split}
& W_{S} (Q^2) = w_1 \,, \quad W_{A_1}(Q^2) = w_2 \,,\quad W_{A_2}(Q^2) = w_1 + w_3 - \frac{w_2}{2 m_p^2} \left(m_p^2-Q^2-m_\pi^2\right) \,, \quad \\[2mm]
& W_{A_3}(Q^2) = -w_3 \,,\quad W_{A_4}(Q^2) = -w_4 + \frac{w_3}{2 m_p^2} \left(m_p^2 - Q^2 - m_\pi^2 \right) \,,\\[2mm]
& W_{T_1}(Q^2) = w_4 \,,\quad W_{T_2}(Q^2) = w_3 \,,\quad W_{T_3}(Q^2)  = w_2 \,,
\end{split}
\label{eq:WXhad}
\eea
with $m_\pi \simeq 135 \, {\rm MeV}$ denoting the mass of the neutral pion.

\section{Structure of LCSRs}
\label{sec:anatomy}

The derivation of the QCD results for the LCSRs proceeds in full analogy to Section~3 of our earlier work~\cite{Haisch:2021hvj} to which we refer the interested reader for all technical details. 
The~analytic expressions for the QCD correlation functions relevant in the context of this article can be found in Appendix~\ref{app:formulaeLCSRs}. Rather than repeating the necessary steps to derive them, let us discuss the structure of the $\Pi_{\alpha}^{\text{QCD}}$  functions. A striking feature of the results for the QCD correlation functions  is that
\begin{equation} \label{eq:T3strike}
\Pi_{T_3}^{\text{QCD}} = 0 \,, 
\end{equation}
at the lowest order in the twist expansion. The first non-zero correction to the QCD correlation function $\Pi_{T_3}^{\text{QCD}}$ schematically takes the form 
\begin{equation} \label{eq:twist4}
\Pi_{T_3}^{\text{QCD}} \sim \braket{\bar{q} q} \cdot \braket{\pi^0 | \, \bar{q}(0) \gamma^\mu \,i g_s \widetilde{G}^{\alpha \beta}(u x) \tau^3 q(x) \,| 0} \,. 
\end{equation}
Here $\braket{\bar{q} q} $ is the quark condensate, $q \equiv (u \; d)^T$, $\tau^3 \equiv   \sigma^3/2$ with $\sigma^3 = \text{diag} \left (1,-1 \right )$ the third Pauli matrix, $g_s$ is the QCD coupling constant and we have defined $\widetilde{G}^{\alpha\beta} \equiv \epsilon^{\alpha \beta \mu \nu} G_{\mu \nu}^A T^A/2$ with $G_{\mu \nu}^A$ the QCD field strength tensor and $T^A$ the $SU(3)$ colour generators. The contribution~(\ref{eq:twist4}) corresponds to a three-particle pion distribution amplitude (DA) of twist~4 with the fields evaluated at $0$, $ux$ and $x$. See~for instance~\cite{Ball:1998je,Ball:2004ye} for details. Being of higher twist the correction~(\ref{eq:twist4}) is expected to  be  small compared to the values predicted for the form factor $w_2$ by the LCSRs for~$\Pi_{A_1}$ and~$\Pi_{A_2}$ --- cf.~(\ref{eq:WXhad}).  This implies that there has to be a cancellation among the contributions of the ground state and that of heavier states in the hadronic sum leading to~$\Pi_{T_3}^{\text{had}} \simeq 0$. Similar cancellations are also observed in certain QCD sum rules for the nucleon mass~\cite{Ioffe:1981kw,Ioffe:1982ce}. In this case, for a specific choice of the nucleon interpolating current, one of the sum rules starts at higher order in the operator production expansion~(OPE), which numerically yields a very small value for the QCD side of the sum rule.  In this example, on the hadronic side  excitations of the nucleon with even and odd parity contribute with opposite signs leading to a cancellation.  In the case of~(\ref{eq:T3strike})  the contributions of excited states are not sign-definite, but in principle cancellations may occur if the contributions from heavier states are sizeable. Sum rules that have this feature cannot be used to extract the form factors related to the ground state because the corrections of excited states are just as important as the formally leading ground-state contributions. The~LCSR for the correlation function~$\Pi_{T_3}$ is thus disregarded in our work. 

Convergence criteria are now applied to the remaining LCSRs in order to determine the Borel window for each $\Pi_{\alpha}$. Notice that compared to the correlator studied for GUT-like proton decay in~\cite{Haisch:2021hvj} the hadronic representation of the correlation function~\eqref{eq:corhadexpl} comprises a larger number of independent Lorentz structures.  This feature leads to simpler analytic LCSR expressions for the correlation  functions $\Pi_\alpha$, but it also renders the numerical impact of the dimension-five condensate $\braket{\bar{q}g_s G\cdot \sigma q}$ with $G \cdot \sigma \equiv G_{\mu\nu} \hspace{0.25mm} \sigma^{\mu \nu}$ larger than in the GUT~case. As a result  the LCSRs analysed below will have larger uncertainties than those that have been studied in~\cite{Haisch:2021hvj}.  In the following, we will use the LCSRs for $\Pi_S$, $\Pi_{A_1}$, $\Pi_{A_2}$ and $\Pi_{T_1}$ to extract the form factors $w_n$  because they are the most well behaved with regard to the power suppression of higher-dimensional condensates  and the dominance of the ground-state contributions. We add that the LCSR  for $\Pi_{A_4}$ also fulfils the convergence criteria but  one would need to combine it with the result for $\Pi_{T_1}$ to extract the form factor~$w_3$ and it turns out that the Borel windows of these two LCRSs do not overlap. 

\begin{figure}[t]
\centering
\includegraphics[width=\textwidth]{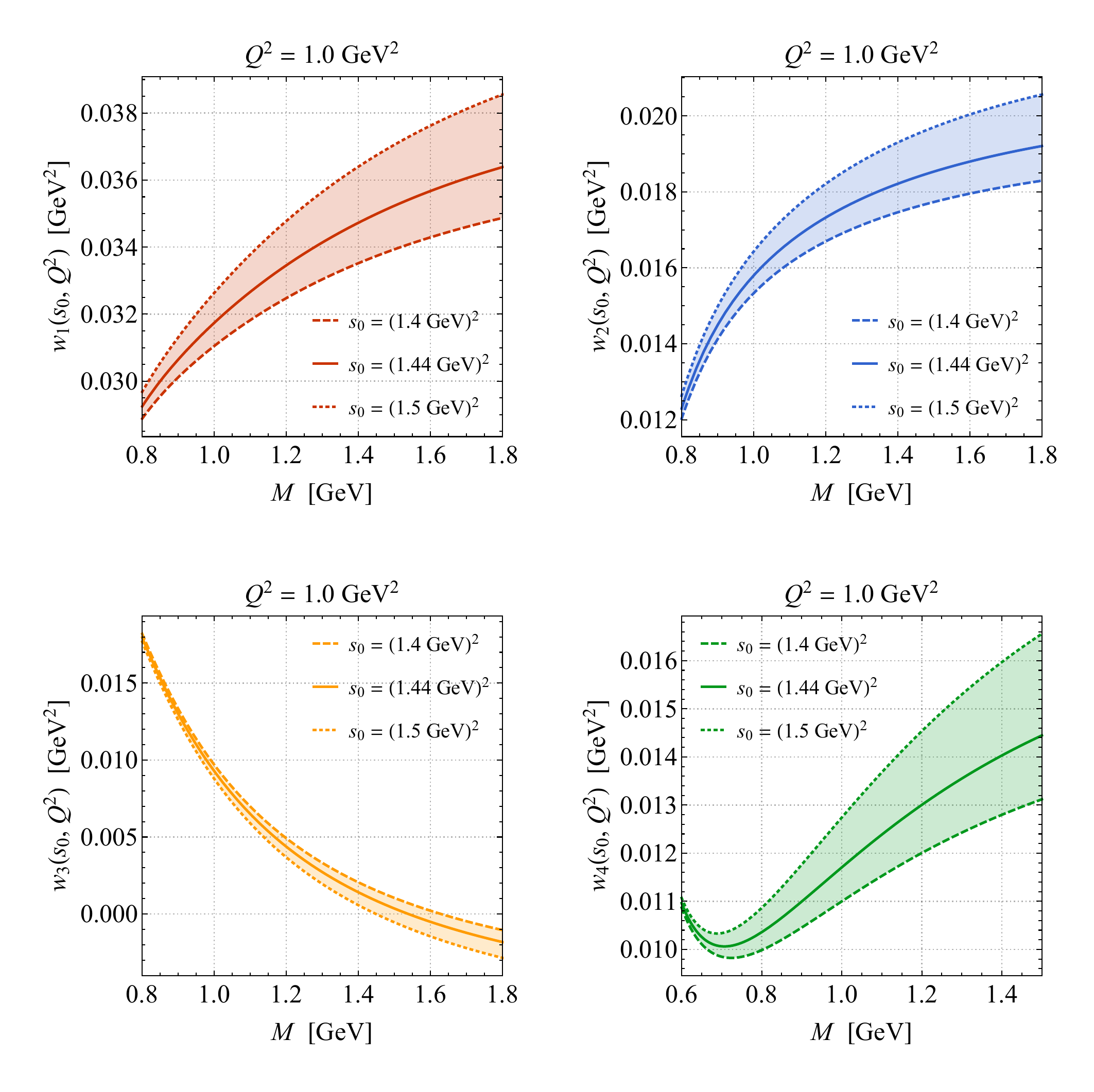}
\vspace{-10mm}
\caption{Form factors  $w_n(s_0, Q^2)$ as a function of the Borel mass $M$ for three different values of the continuum threshold $s_0$. All plots show the results at $Q^2 = 1 \, {\rm GeV}^2$.}
\label{fig:borel}
\end{figure}

In the case of the LCSRs for $\Pi_S$ and $\Pi_{A_2}$ we find the window $1.1 \text{ GeV} \lesssim M \lesssim 1.5 \text{ GeV}$ with  $M$  the Borel mass, while for~$\Pi_{T_1}$ we obtain   $0.7 \text{ GeV} \lesssim M \lesssim 1.1 \text{ GeV}$. In all three cases the Borel analysis has been restricted to~$0.6 \text{ GeV}^2 \lesssim Q^2 \lesssim 2.5 \text{ GeV}^2$. The lower limits are obtained by demanding that the  mixed condensate $\braket{\bar{q}g_s G\cdot \sigma q}$  does not account for more than~$50\%$ of the total QCD result and as an absolute minimum of the Borel mass we choose $0.7 \, {\rm GeV.}$ The upper limits arise from the requirement that heavier states constitute less than~$50\%$ of the total dispersion integrals~\eqref{eq:haddisp} and that the Borel mass should not considerably exceed the continuum threshold~$s_0 =1.44 \text{ GeV}^2$. The latter is chosen as the square of the mass of the Roper resonance~\cite{Zyla:2020zbs} which is the lightest excitation in the nucleon spectrum. The Borel transformation ensures that heavier states with a mass~$m_{N^\prime}$ are exponentially suppressed by a factor of~$\exp \left (-(m_{N^\prime}^2-m_p^2)/M^2 \right)$, but the dispersion integral over heavier states, which starts at~$s_0$, can be modelled as an integral over the QCD result assuming  quark-hadron duality~\cite{Poggio:1975af,Shifman:2000jv}.  

In order to determine all four form factors $w_n$ one also needs to evaluate~$\Pi_{A_1}$, where the power suppression turns out to be less effective than in the other cases. Therefore the use of this LCSR is restricted to~$Q^2 \gtrsim 0.9 \, {\rm  GeV}^{2}$ because the power suppression becomes more effective for larger values of $Q^2$. For Borel masses of $1.1 \text{ GeV} \lesssim M \lesssim 1.5 \text{ GeV}$ the relative contribution of the dimension-five condensate  $\braket{\bar{q}g_s G\cdot \sigma q}$  to the LCSR lies between $60\%$ and $100\%$.  The form factor $w_2$ can therefore only be estimated within systematic uncertainties of~the order of~$100\%$. The result of $w_2$ also enters the prediction for $w_3$ through the LCSR for $\Pi_{A_2}$ but the contribution is suppressed by a kinematical factor of about~$0.1$~for~$Q^2 \simeq 1 \text{ GeV}^2$~$\big($cf.~\eqref{eq:WXhad}$\big)$. As a result the uncertainties plaguing $w_2$ represent only a subleading part of the uncertainty in  $w_3$. In fact, it turns out that the  differential decay width of $p\to e^+ \pi^0 G$ receives the dominant contributions from the form factors $w_1$ and~$w_3$, meaning that the uncertainty due to $w_2$ plays only a minor role in the proton decay phenomenology in the GRSMEFT. 

If one could resum the expansion of the QCD side to all orders, the dependence on the Borel mass $M$ of the form factors would vanish. Truncating the expansion at some finite order leaves a residual dependence on this parameter, but ideally the results for the form factors $w_n$  do not depend too strongly on the exact choice of this scale. Moreover, observables should not depend on the choice of the effective continuum threshold~$s_0$ which enters the LCSRs because like in~\cite{Haisch:2021hvj} we assume quark-hadron duality~\cite{Poggio:1975af,Shifman:2000jv}. These two scales are therefore referred to as unphysical~parameters in the following. Figure~\ref{fig:borel}~displays the dependence of the form factors  on the Borel mass $M$ and the continuum threshold~$s_0$, where $s_0$ is varied between $(1.4 \, {\rm GeV})^2$ and $(1.5 \, {\rm GeV})^2$. The broader the obtained band the stronger is the dependence of the form factor on $s_0$, and the steeper the curves the stronger is the dependence on $M$. By varying the Borel mass within the corresponding window and the continuum threshold between $(1.4 \, {\rm GeV})^2$ and $(1.5 \, {\rm GeV})^2$ one can  obtain an uncertainty estimate for the relevant form factor $w_n$.

\section{Numerical analysis}
\label{sec:numerics} 

\begin{figure}[t]
\centering
\includegraphics[width=\textwidth]{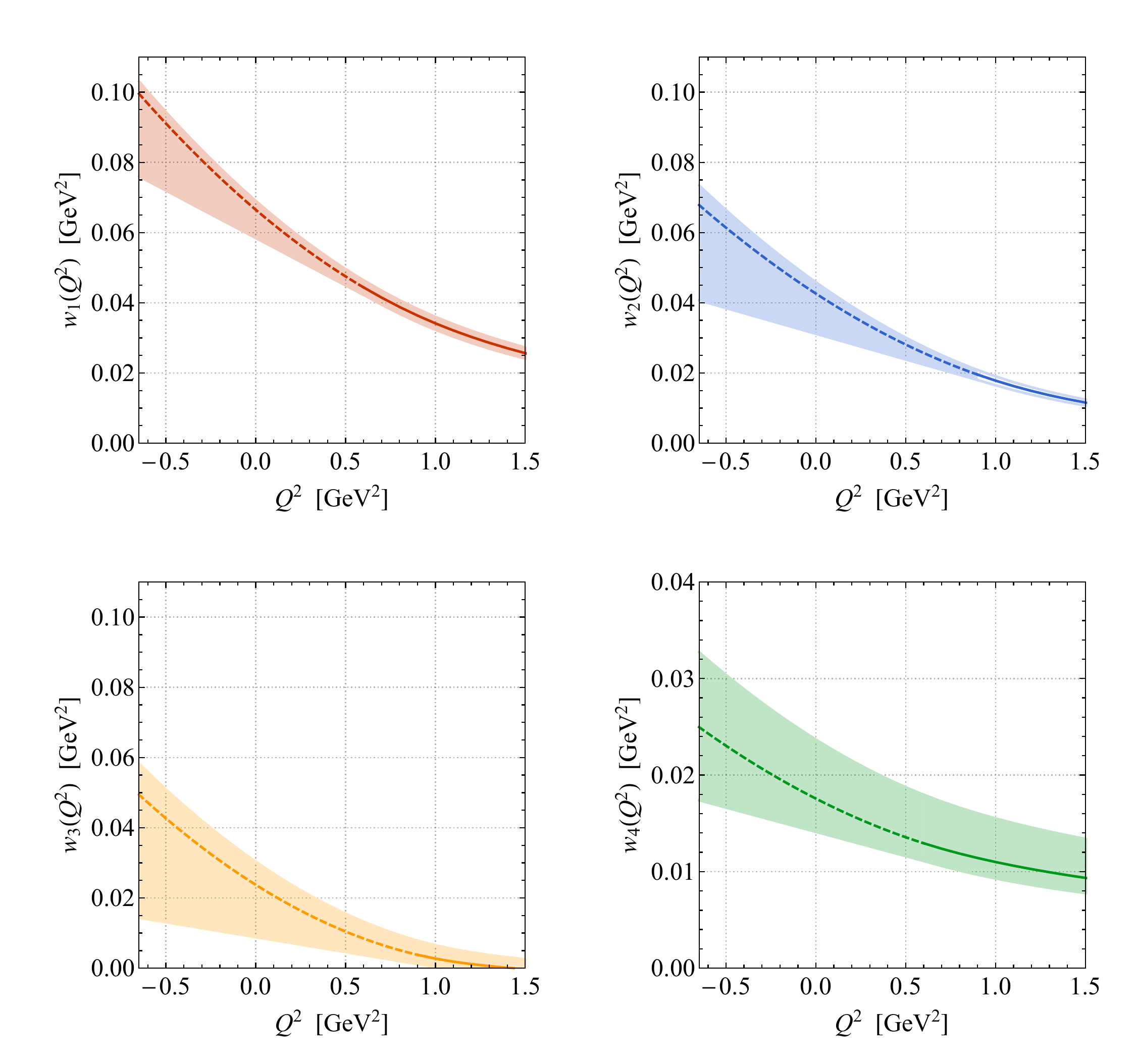}
\vspace{-5mm}
\caption{Form factors $w_n (Q^2)$ as a function of $Q^2$. The coloured curves and bands correspond to the central values and uncertainties of the LCSRs. In the case of $w_1(Q^2)$ and $w_4 (Q^2)$ $\big($$w_2(Q^2)$ and $w_3 (Q^2)$$\big)$ the predictions for $0.6 \, {\rm GeV}^2 \leq Q^2 \leq 1.5 \, {\rm GeV}^2$ ($0.9 \, {\rm GeV}^2 \leq Q^2 \leq 1.5 \, {\rm GeV}^2$) are obtained by a direct calculation~(solid lines), while the predictions for $Q^2 \leq 0.6 \, {\rm GeV}^2$ ($Q^2 \leq 0.9 \, {\rm GeV}^2$) are obtained by an extrapolation~(dashed lines). Consult the main text for further explanations.}
\label{fig:ffplot}
\end{figure}

Using the numerical input and the definitions of the pion DAs of Appendix~\ref{app:input} we obtain the results for the form factors~$w_n$ shown in Figure~\ref{fig:ffplot}. The displayed central values of~$w_n$ correspond to  $M = 1.3 \, {\rm  GeV}$ for $\Pi_S$, $\Pi_{A_1}$ and $\Pi_{A_2}$, while in the case of  $\Pi_{T_1}$ we use $M = 0.9 \, {\rm  GeV}$. All central predictions employ $s_0 = (1.44 \, {\rm GeV})^2$. The total theoretical uncertainties receive contributions from variations of $M$ and $s_0$ as described above but also from variations of the numerical input parameters within their uncertainties (cf.~Appendix~\ref{app:input}). Each parameter is varied independently while the remaining parameters are kept fixed at their central values. The total uncertainty is then obtained by adding individual uncertainties in quadrature. The values of the form factors $w_1(Q^2)$ and $w_4(Q^2)$ are computed with the help of the LCSRs for $Q^2 \geq 0.6 \, {\rm GeV}^2$ while for $Q^2 \leq 0.6 \, {\rm GeV}^2$ the values are obtained by a naive extrapolation. Similarly, the form factors $w_2(Q^2)$ and $w_3(Q^2)$ are predicted by the LCSRs for $Q^2 \geq 0.9 \, {\rm GeV}^2$ and by the extrapolation for $Q^2 \leq 0.9 \, {\rm GeV}^2$.  A linear and a quadratic function in $Q^2$ is taken  to extrapolate  the form factors which are then fitted to the results of the values obtained from the LCSRs in the vicinity of $Q^2 = 0.6 \, {\rm GeV}^2$ for $w_1(Q^2)$ and $w_4(Q^2)$ and $Q^2 = 0.9 \, {\rm GeV}^2$ for $w_2(Q^2)$ and $w_3(Q^2)$. For a given form factor the quadratic fit is chosen to obtain the central value for $w_n$, while the maximum and minimum of all extrapolations determine the uncertainty band. We remark that the same fitting approach has been successfully used in the LCSR calculation~\cite{Haisch:2021hvj} to reproduce the results from LQCD in the case of a GUT-like proton decay. 

The fit formulas for the form factors $w_n$ that we obtain in the physical region,~i.e.~in the four-momentum range $-0.65 \, {\rm GeV}^2 \simeq -(m_p - m_{\pi})^2 \leq Q^2 \leq  0$, take the following form
\beq \label{eq:w1}
w_1 (Q^2) = \left \{ \begin{matrix} \left [ 0.012  \left ( \displaystyle \frac{Q^2}{{\rm GeV}^2} \right )^2-0.045  \left ( \displaystyle  \frac{Q^2}{{\rm GeV}^2} \right )  +0.069  \right ]  {\rm GeV}^2 \,,  \\[5mm] 
\left [ 0.011 \left ( \displaystyle \frac{Q^2}{{\rm GeV}^2} \right )^2  -0.044  \left ( \displaystyle  \frac{Q^2}{{\rm GeV}^2} \right ) +0.066 \right ] {\rm GeV}^2 \,, \\[5mm] 
\left [ 0.058 -0.027  \left ( \displaystyle \frac{Q^2}{{\rm GeV}^2} \right )  \right ] {\rm GeV}^2 \,, \end{matrix}  \right . 
\eeq
\beq  \label{eq:w2}
w_2 (Q^2) = \left \{ \begin{matrix} \left [ 0.009  \left ( \displaystyle \frac{Q^2}{{\rm GeV}^2} \right )^2-0.036  \left ( \displaystyle  \frac{Q^2}{{\rm GeV}^2} \right )  +0.046  \right ]  {\rm GeV}^2 \,,  \\[5mm] 
\left [ 0.008 \left ( \displaystyle \frac{Q^2}{{\rm GeV}^2} \right )^2  -0.033  \left ( \displaystyle  \frac{Q^2}{{\rm GeV}^2} \right ) +0.043 \right ] {\rm GeV}^2 \,, \\[5mm] 
\left [ 0.031 -0.015  \left ( \displaystyle \frac{Q^2}{{\rm GeV}^2} \right )  \right ] {\rm GeV}^2 \,, \end{matrix}  \right . 
\eeq
\beq  \label{eq:w3}
w_3 (Q^2) = \left \{ \begin{matrix} \left [ 0.012  \left ( \displaystyle \frac{Q^2}{{\rm GeV}^2} \right )^2-0.035  \left ( \displaystyle  \frac{Q^2}{{\rm GeV}^2} \right )  +0.030  \right ]  {\rm GeV}^2 \,,  \\[5mm] 
\left [ 0.011 \left ( \displaystyle \frac{Q^2}{{\rm GeV}^2} \right )^2  -0.032  \left ( \displaystyle  \frac{Q^2}{{\rm GeV}^2} \right ) +0.024 \right ] {\rm GeV}^2 \,, \\[5mm] 
\left [ 0.009 -0.009  \left ( \displaystyle \frac{Q^2}{{\rm GeV}^2} \right )  \right ] {\rm GeV}^2 \,, \end{matrix}  \right . 
\eeq
\beq  \label{eq:w4}
w_4 (Q^2) = \left \{ \begin{matrix} \left [ 0.004  \left ( \displaystyle \frac{Q^2}{{\rm GeV}^2} \right )^2-0.012  \left ( \displaystyle  \frac{Q^2}{{\rm GeV}^2} \right )  +0.024  \right ]  {\rm GeV}^2 \,,  \\[5mm] 
\left [ 0.003 \left ( \displaystyle \frac{Q^2}{{\rm GeV}^2} \right )^2  -0.009  \left ( \displaystyle  \frac{Q^2}{{\rm GeV}^2} \right ) +0.018 \right ] {\rm GeV}^2 \,, \\[5mm] 
\left [ 0.014 -0.005  \left ( \displaystyle \frac{Q^2}{{\rm GeV}^2} \right )  \right ] {\rm GeV}^2 \,. \end{matrix}  \right . 
\eeq
Here the upper (lower) line in each formula corresponds to the upper (lower) border of the corresponding envelope shown in  Figure~\ref{fig:ffplot}, while the middle line represents the central value of our LCSR form factor prediction. 

Let us also discuss alternative approaches for extrapolating the LCSR results for the form factors to the physical regime,~i.e.~to negative~$Q^2$. A physically well-motivated approach is to model the lowest-lying resonance(s) in the $Q^2$-spectrum explicitly by poles in the complex $t\equiv -Q^2$ plane and capture modifications to the spectrum due to other contributions by an additional series expansion in a new variable~$z(t)$~---~see~for example~\cite{Boyd:1994tt,Boyd:1997qw,Becher:2005bg,Arnesen:2005ez,Bharucha:2010im} for details on this so-called $z$-expansion.  This parametrisation is often used to take into account resonances that lie below the threshold of the two-particle branch cut which in our case is located at $t=t_{\rm th}\equiv (m_p+m_\pi)^2$. However, the lowest-lying resonance is heavier than~$t_{\rm th}$ in the case at hand so it is not clear whether the form factors are dominated by a pole contribution close to the upper limit of the allowed three-body kinematics,~i.e.~$t= (m_p - m_\pi)^2$, or by the continuum two-particle contribution. Employing the $z$-expansion up to the second order including a single pole at the mass  $m_{\Delta^+} = 1.232 \, {\rm GeV}$~\cite{Zyla:2020zbs} of the $\Delta^+$ resonance leads to a steeper increase in magnitude of the form factors $w_n$ at small $Q^2$ and in some cases to significantly  narrower uncertainty bands. This approach would therefore lead to larger predictions for the form factors $w_n$. In another approach used in~\cite{Ball:2004ye,Becirevic:1999kt}, all contributions but the lowest-lying resonance are modelled by an effective pole at higher mass  which is more flexible than a single-pole fit. For the form factors~$w_n$ this procedure however generates unphysical singularities in the regime of physical~$Q^2$, rendering it unsuitable. From the above, we conclude that our power-expansion approach~\eqref{eq:w1}~to~\eqref{eq:w4} yields a more conservative estimate of the form factors $w_n$ at small~$Q^2$ than the~$z$-expansion including a single pole and is more reliable than the effective two-pole fit. It is therefore preferred with respect to extrapolations featuring explicit poles in the $Q^2$-spectrum for the sensitivity studies of semi-leptonic proton decay searches that are discussed in the following section.

We add that the form factors $w_n$ are related to the off-shell form factors of the decomposition of the more general matrix element $\braket{\pi^0 | \epsilon^{abc} \hspace{0.125mm} d_a^\alpha  \hspace{0.125mm}  u_b^\beta  \hspace{0.125mm}  u_c^\gamma | p}$~\cite{Pire:2005ax,Pire:2021hbl}, where $\alpha$, $\beta$ and $\gamma$ are Dirac indices.  Certain combinations of the form factors $w_n$ therefore yield the form factors $W^k_{RR}$ with $k=0,1$ that are relevant for GUT-like proton decay. In~Appendix~\ref{app:check} we show that using the results~(\ref{eq:w1}) to~(\ref{eq:w4}) allows one to reproduce the physical values of the form factors~$W^k_{RR}$  calculated in our previous work~\cite{Haisch:2021hvj} within uncertainties. This gives us confidence that the naive extrapolation used to obtain the above expressions for~$w_n$  sufficiently approximates the true scaling in the relevant four-momentum regime. 

\section{Proton decay phenomenology}
\label{sec:pheno}

With the help of~the expressions~(\ref{eq:w1}) to~(\ref{eq:w4})  the $p \to e^+ \pi^0 G$ decay amplitude~\eqref{eq:pdecampl} can now be calculated. One first notices that after making use of the on-shell conditions for the graviton~$\big($cf.~\eqref{eq:transverse}$\big)$ the contribution of $w_4$ vanishes. This feature can be understood by means of the  soft pion theorem~\cite{Nambu:1997wa,AdlerDashen68}. In fact, in the soft pion limit and recalling that $q =p_p-p_\pi$  one finds that all terms but the contribution of $w_4$ vanish in the hadronic tensor: 
\begin{equation} \label{eq:Hw4}
\lim_{p_\pi \to 0} H^{\mu\nu} (p_p, q)  = H^{\mu\nu} (p_p, p_p) = i w_4 P_R \sigma^{\mu\nu} u_p(p_p) \,.
\end{equation}
However, in the   limit $p_\pi \to 0$ the pion can be removed from the decay amplitude giving rise to the following relation 
\begin{equation}
\begin{split}
\lim_{p_\pi \rightarrow 0} \braket{e^+ G \pi^0(p_\pi) | \mathcal{L}_{\slashed{B}}^{(8)}(0) |p} =& -\frac{i}{\sqrt{2}f_\pi} \braket{e^+ G|\mathcal{L}_{\slashed{B}}^{(8)}(0) |p} \\[2mm]
& + \lim_{p_\pi \rightarrow 0} \frac{\sqrt{2}}{f_\pi} p_\pi^\mu \int\! d^4x \, e^{ip_\pi x} \braket{e^+ G| \hspace{0.125mm}T\hspace{0.125mm} \big[ J^A_\mu (x)\mathcal{L}_{\slashed{B}}^{(8)}(0)\big]|p} \,,
\label{eq:softpion}
\end{split}
\end{equation}
where $J^A_\mu (x) \equiv \left [ \bar{u}(x) \gamma_\mu \gamma_5 u (x) -  \bar{d}(x) \gamma_\mu \gamma_5 d (x) \right ]/2$ denotes the axial current and the pion field is related to axial current by $\pi^0(x) = \sqrt{2} \, \partial^\mu J^A_\mu (x) / (f_\pi m_\pi^2)$. The second term in~\eqref{eq:softpion} vanishes unless there are additional poles in the soft pion limit. Such poles occur if the pion is attached to one of the external lines~\cite{AdlerDashen68} in the $p \rightarrow e^+ G$ amplitude, which is formally described by inserting a complete set of intermediate states between the operators in the time-ordered product. The pion however can only couple to the external proton line, so pole contributions arise only when the pion is emitted from the incoming proton. This type of correction thus leads again  to the matrix element of the $p\rightarrow e^+ G$ transition, which however satisfies~$\braket{e^+ G|\mathcal{L}_{\slashed{B}}^{(8)}(0) |p}=0$, because the transition is forbidden by angular momentum conservation. As a result the right-hand side of~\eqref{eq:softpion} vanishes identically:
\begin{equation}
\lim_{p_\pi \rightarrow 0} \braket{e^+ G \pi^0(p_\pi) | \mathcal{L}_{\slashed{B}}^{(8)}(0) |p} = 0 \,.
\end{equation}
Since the form factor $w_4$ itself is non-vanishing it then follows that the associated Lorentz structure~(\ref{eq:Hw4}) does not contribute to the proton decay channel $p \rightarrow e^+ \pi^0   G$ at all.

By squaring the amplitude, summing over spins and polarisations and calculating the phase space integrals the differential decay width can be computed. Note that the transversality of the graviton $\big($see~\eqref{eq:transverse}$\big)$ has been used to drop unphysical contributions which violates the gauge symmetry of gravity in the weak field limit. The gauge symmetry ensures that negative-norm states cancel out in the sum over polarisations. Therefore the sum has to be constrained to physical polarisations only by employing~\cite{vanDam:1970vg}
\begin{equation} \label{eq:polsum}
\sum_\lambda \varepsilon^\ast_{\alpha \beta}(p,\lambda) \hspace{0.5mm}  \varepsilon_{\gamma\delta}(p,\lambda) = \frac{1}{2} \left(\eta^\prime_{\alpha\delta} \eta^\prime_{\beta \gamma} + \eta^\prime_{\alpha\gamma}\eta^\prime_{\beta\delta} - \eta^\prime_{\alpha \beta} \eta^\prime_{\gamma \delta} \right) \,,
\end{equation}
with
\beq
 \eta^\prime_{\mu \nu} \equiv \eta_{\mu\nu} - \frac{\bar{p}_\mu p_\nu+p_\mu \bar{p}_\nu}{p \cdot \bar{p}} \,,
\eeq
and $\bar{p} \equiv (p^0,-\vec{p})$ such that $\bar{p}^2=0$.

Neglecting the mass of the  positron but keeping the mass of the neutral pion, the $p \to e^+ \pi^0 G$ rate corresponding to the  fiducial region of the three-particle phase space defined by an upper cut on the graviton energy can be written as  
\bea \label{eq:Gammaprotondiff}
\begin{split}
\Gamma \left ( p \to e^+ \pi^0 G \right )_{{\rm fid}} & =  \frac{m_p^7 \hspace{0.5mm} |c_{\slashed{B}}|^2}{128 \hspace{0.25mm}   \pi^3   \overbar{M}_{\hspace*{-1.5pt} {P}}^2} \, \int_{0}^{y_{\rm fid}}  \! dy    \int_{z_0}^{z_1} \! dz \left (x_{\pi} + y + z -y \hspace{0.25mm} z -1 \right) \left (x_{\pi}+y+z-1 \right )^2  \hspace{6mm} \\[2mm]
& \phantom{xx} \times \left [  l_1 \left ( w_1 + w_3 \right )^2 +  l_2 \hspace{0.25mm}  w_2^2 + l_3 \hspace{0.25mm}  w_2 \left (w_1 + w_3 \right )  \right ]  \,.
\end{split}
\eea
Here we have defined $y \equiv 2  E_G/m_p$ and $z \equiv 2  E_e/m_p$ with $E_G$ ($E_e$) the graviton (positron) energy in the rest frame of the proton, $x_\pi\equiv m_\pi^2/m_p^2$, the boundaries for the integral over $z$ are given by 
\beq
z_0 = 1 - x_\pi - y \,, \qquad z_1 = \frac{1 - x_\pi - y}{1 - y} \,, 
\eeq
and 
\beq
l_1 = - 4 \hspace{0.25mm} y \,, \quad 
l_2  = y \, \big [ 4 \hspace{0.25mm}  x_{\pi} - \left ( y + z - 2 \right )^2 \big ]  \,, \quad 
l_3  = - 4 \, \big [ 2 \hspace{0.25mm} x_{\pi} - (y + z)  \left ( y-2 \right ) -2 \big ]  \,.
\eeq
When expressed through  the integration variables of~(\ref{eq:Gammaprotondiff}) the scale $Q^2$ that enters the form factors $w_n$ finally takes  the following form 
\beq
Q^2 = -q^2 = -m_p^2 \, \big ( x_\pi + y + z - 1 \big  )\,. 
\eeq

The GRSMEFT  proton decay mode $p \to e^+ \pi^0 G$ experimentally leads to events that contain a positron, two photons arising from the decay of the neutral pion and missing energy~($E^{\rm miss}$) because the graviton escapes the detector undetected. Such a signature has to our knowledge not been searched for directly in experiments that study the possible decays of the proton. As we will show, existing searches  that are however sensitive to  $p \to e^+ \pi^0 G$  are the inclusive search $p \to e^+  X$ with $X$ an arbitrary final state and the exclusive search for $p \to e^+  \pi^0$. The total inclusive rate $p \to e^+  X$ can be obtained by employing $y_{\rm fid} = 1 - x_\pi$ in~(\ref{eq:Gammaprotondiff}). Numerically, we find that 
\beq \label{eq:widthproton}
\Gamma \left ( p \to e^+ \pi^0 G \right ) =  \frac{m_p^7 \hspace{0.5mm} \Lambda_p^4 \hspace{0.5mm} |c_{\slashed{B}}|^2}{256 \hspace{0.25mm}   \pi^3  \overbar{M}_{\hspace*{-1.5pt} {P}}^2} \,, \qquad \Lambda_p = \left ( 99 \pm 13 \right ) {\rm MeV} \,.
\eeq
Here we have introduced the hadronic parameter $\Lambda_p$ and the normalisation factor $1/(256  \hspace{0.25mm}\pi^3 )$  takes into account the phase-space suppression for a three-body decay. The uncertainty on~$\Lambda_p$ is obtained by calculating the minimal and maximal rate that can be achieved by considering all possible combinations of the form factor parameterisations~(\ref{eq:w1})~to~(\ref{eq:w4}). Notice that since~$\Lambda_p$ appears in~(\ref{eq:widthproton}) to the fourth power the LCSR prediction for $\Gamma \left ( p \to e^+ \pi^0 G \right )$ has an uncertainty of order 50\%. The~theory uncertainties  of $\Gamma \left ( p \to e^+ \pi^0 G \right )$ are therefore significantly larger than those that plague the GUT predictions for $\Gamma \left ( p \to e^+ \pi^0  \right )$~ obtained in both LQCD~\cite{Gavela:1988cp,Aoki:1999tw,Tsutsui:2004qc,Aoki:2006ib,Braun:2008ur,Aoki:2013yxa,Aoki:2017puj,Yoo:2018fyn} and LCSRs \cite{Haisch:2021hvj}.  

\begin{figure}[t!]
\centering
\includegraphics[scale=.8]{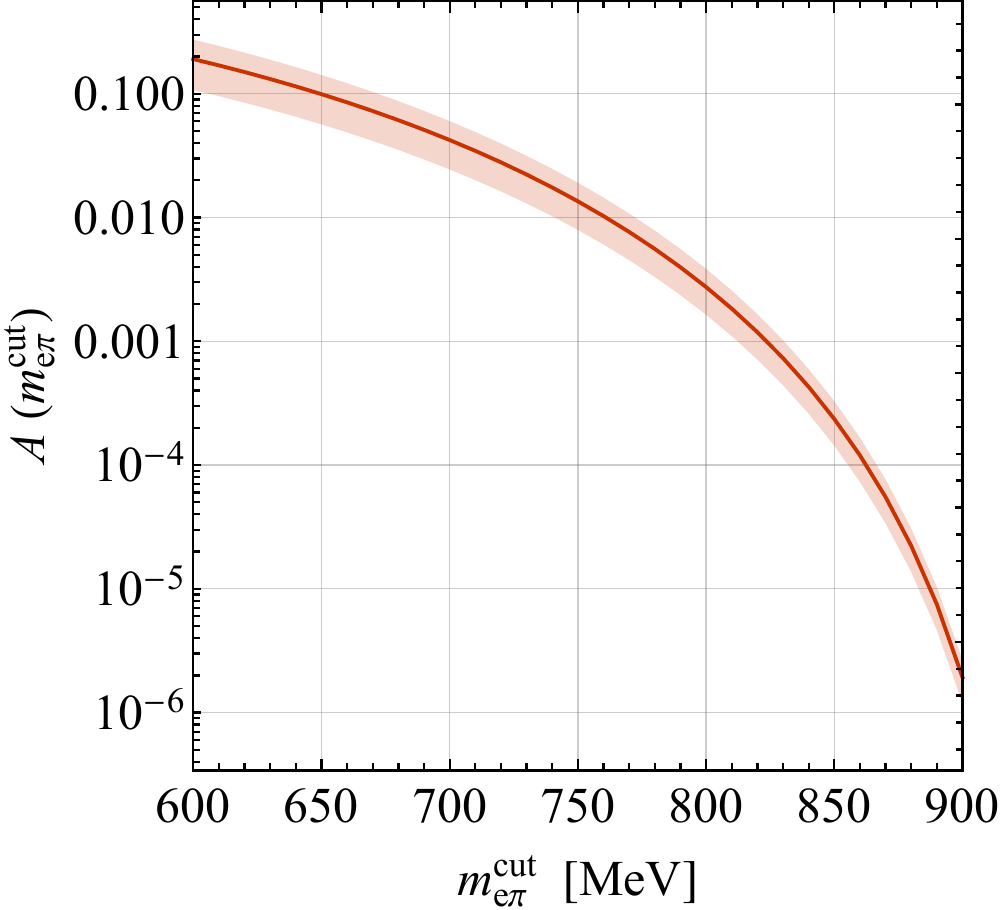}
\vspace{0mm}
\caption{\label{fig:acceptance} Acceptance~(\ref{eq:acceptance}) as a function of the cut $m_{e \pi}>m_{e \pi}^{\rm cut}$ on the invariant mass  of the $e^+ \pi^0$ system. The red curve indicates the central prediction, while the red band illustrates the maximal variations that result from considering all possible combinations of the form factor parameterisations~(\ref{eq:w1})~to~(\ref{eq:w4}) in the fiducial decay width keeping the inclusive decay width fixed at the central value. See text for further details. }
\end{figure}

Searches for the two-body decay mode $p \to e^+  \pi^0$ can also be used to set a bound on the GRSMEFT interaction~(\ref{eq:dim8}), because the cuts that experiments such as SK impose do not fully eliminate the contributions that arise from  $p \to e^+  \pi^0 G$. The relevant requirements in these experiments are selections on the invariant mass $m_{e\pi}$ and the magnitude of the three-momentum $p_{e\pi}$ of the $e^+ \pi^0$ system. In the rest frame of the proton these quantities can be expressed in terms of the graviton energy  as $m_{e\pi} = \sqrt{m_ p \left (m_p - 2 E_G \right )}$ and $p_{e\pi} = E_G$.     In the latest SK search~\cite{Super-Kamiokande:2020wjk} the definition of the signal region involves the requirements $m_{e\pi} > 800 \, {\rm MeV}$ and $p_{e \pi} < 250 \, {\rm MeV}$, meaning that all $p \to e^+ \pi^0 G$ events that satisfy the~$m_{e\pi}$ selection also pass the~$p_{e \pi}$ cut. To quantify by how much a lower cut $m_{e \pi}> m_{e \pi}^{\rm cut} $ reduces the observed $p \to e^+ \pi^0 G$ rate we introduce the acceptance 
\beq \label{eq:acceptance}
A \left (m_{e \pi}^{\rm cut} \right )  = \frac{\Gamma \left ( p \to e^+ \pi^0 G \right )_{\rm fid}}{\Gamma \left ( p \to e^+ \pi^0 G \right )}  \,,
\eeq
where $\Gamma \left ( p \to e^+ \pi^0 G \right )_{\rm fid}$ is the fiducial width~(\ref{eq:Gammaprotondiff}) evaluated setting $y_{\rm fid} = 1 - \big ( m_{e \pi}^{{\rm cut}}/m_p \big)^2$ and $\Gamma \left ( p \to e^+ \pi^0 G \right )$ is the total inclusive width given in~(\ref{eq:widthproton}). In Figure~\ref{fig:acceptance} we show our predictions for $A \left (m_{e \pi}^{\rm cut}  \right )$ in the range of $m_{e \pi}^{\rm cut}$ that is relevant for searches for the two-body proton decay mode $p \to e^+ \pi^0$ at existing and next-generation water Cherenkov detectors.

We are now in a position to derive bounds on the Wilson coefficient $c_{\slashed{B}}$ that multiplies the dimension-eight operator  in~(\ref{eq:dim8}).  We begin with the inclusive   $p \to e^+  X$ decay. The~currently best proton lifetime limit from $p \to e^+  X$ is unfortunately more than 40~years old. It reads~\cite{Learned:1979gp}
\beq \label{eq:taup1979}
\tau_p \left ( p \to e^+  X \right )  > 0.6 \cdot 10^{30} \, {\rm yr} \,, 
\eeq
and together with~(\ref{eq:widthproton})   leads to 
\beq \label{eq:c6bound}
|c_{\slashed{B}}| < (104 \, {\rm GeV} )^{-4} \,,
\eeq
at the 90\%~CL. It has been pointed out in~\cite{Heeck:2019kgr} that with existing  data from water Cherenkov detectors  it should be possible to set a significantly better limit on $p \to e^+  X$ compared to the proton lifetime limit reported in~(\ref{eq:taup1979}).  An~estimate of such an improved  limit can be obtained from the limit of $1.7 \cdot 10^{32} \, {\rm yr}$ at~90\%~CL on the $p \to e^+ + E^{\rm miss}$  channel~\cite{Super-Kamiokande:2014pqx} since the latter decay bears close resemblance to the inclusive $p\to e^+ X$ mode. In fact, the authors of~\cite{Heeck:2019kgr} estimated that with the available SK data it should be possible to improve~(\ref{eq:taup1979}) by around two orders of magnitude.  Note that  a factor of 100 improvement on $\tau_p \left ( p \to e^+  X \right )$ would push  the bound~(\ref{eq:c6bound}) up to  $(186 \, {\rm GeV} )^{-4}$.

In order to determine the  SK sensitivity to the $p \to e^+ \pi^0 G$ signature that derives from the measurement~\cite{Super-Kamiokande:2014pqx}, we  need to compute the acceptance~(\ref{eq:acceptance}) for $m_{e \pi}^{\rm cut} = 800 \, {\rm MeV}$. Using the central value of  the hadronic parameter~$\Lambda_p$ as given in~(\ref{eq:widthproton}) we find 
\beq \label{eq:ASK}
A \left ( 800 \, {\rm MeV} \right ) = 2.7 \cdot 10^{-3} \left ( 1 \pm 0.40 \right ) \,.  
\eeq
The smallness of the acceptance is  compensated by the fact that the 90\%~CL lower limit on the lifetime of the proton in $p \to e^+  \pi^0$ is by more than five orders of magnitude better than~(\ref{eq:taup1979}) since  one has~\cite{Super-Kamiokande:2020wjk} 
\beq \label{eq:SK}
\tau_p \left ( p \to e^+  \pi^0  \right )  > 2.4 \cdot 10^{34} \, {\rm yr} \,.
\eeq
Combining (\ref{eq:widthproton}) for $\Lambda_p = 99 \, {\rm MeV}$ with (\ref{eq:ASK})  and (\ref{eq:SK}) one obtains 
\beq \label{eq:c6boundbetter}
|c_{\slashed{B}}| < (185 \, {\rm GeV} )^{-4} \,, 
\eeq
at the 90\%~CL. Notice that this bound is very close to the limit that has been quoted above by  assuming a factor of 100 improvement  of $\tau_p \left ( p \to e^+  X \right )$ compared to~(\ref{eq:taup1979}) based on the estimate presented in~\cite{Heeck:2019kgr}.

It is also straightforward to estimate the sensitivity of HK to the Wilson coefficient $c_{\slashed{B}}$. Running HK for eight years it should be possible to set the following 90\%~CL bound
\beq \label{eq:HK}
\tau_p \left ( p \to e^+  \pi^0  \right )  > 1.0 \cdot 10^{35} \, {\rm yr} \,.
\eeq
This limit has been obtained in~\cite{Abe:2011ts} by considering the same signal region for $p \to e^+ \pi^0$ as the latest SK search~\cite{Super-Kamiokande:2020wjk}. Consequently, we can again use  (\ref{eq:widthproton}), (\ref{eq:ASK})  and (\ref{eq:HK}) to arrive~at 
\beq \label{eq:c6boundHK}
|c_{\slashed{B}}| < (222 \, {\rm GeV} )^{-4} \,. 
\eeq
This bound on the Wilson coefficient of the dimension-eight baryon-number violating  operator~(\ref{eq:dim8})  is probably the ultimate limit that can be set with next-generation neutrino detectors, because both JUNO and~DUNE are not expected to reach the HK sensitivity to the $p \to e^+ \pi^0$ mode~$\big($cf.~\cite{JUNO:2015zny,DUNE:2015lol}$\big)$. 

\section{Conclusions and outlook}
\label{sec:conclusions}

In our work, we have employed  LCSR techniques to calculate the form factors which parametrise the hadronic matrix elements of semi-leptonic three-body proton  decays with a pion in the final state.  While the presented formalism and the obtained results are general, we have specifically  focused on  the computation of the differential decay rate for the process $p \to e^+ \pi^0 G$. This channel is the dominant proton decay mode in  the GRSMEFT,  since the two-body transition $p \to e^+ G$ is forbidden by angular momentum conservation. Like in our earlier article~\cite{Haisch:2021hvj} our LCSR study includes  the leading contributions in the light-cone expansion, namely the twist-2 and twist-3 pion DAs --- the explicit expressions  can be found in~Appendix~\ref{app:formulaeLCSRs} --- and we have performed a detailed study of the dependence of the obtained LCSRs on both the unphysical (i.e.~the Borel~mass and the continuum threshold) and the physical (i.e.~the condensates and the pion DAs) parameters. In this way we are able  to provide results and estimate uncertainties for the form factors in the kinematical regime where the momentum transfer~$q$ from the proton to the pion is space-like. We have then extrapolated our LCSR results to the physical regime $0 \leq q^2 \leq (m_p - m_{\pi} )^2$ by means of both a linear and quadratic fit, including the spread of predictions in our uncertainty estimates. The resulting uncertainties turn out to be significantly larger than those that plague the hadronic matrix elements that are relevant in the GUT case~\cite{Gavela:1988cp,Aoki:1999tw,Tsutsui:2004qc,Aoki:2006ib,Braun:2008ur,Aoki:2013yxa,Aoki:2017puj,Yoo:2018fyn,Haisch:2021hvj}.

Employing the LCSR results for the form factors we have then studied the sensitivity of existing and next-generation water Cherenkov detectors in looking for the $p \to e^+ \pi^0 G$ signature. To this purpose we have calculated the rate for $p \to e^+ \pi^0 G$ differentially in the energies of the final state particles. We have then derived bounds on the amount of dimension-eight baryon-number violation in the GRSMEFT considering both the inclusive search for $p \to e^+ X$~\cite{Learned:1979gp,Heeck:2019kgr} and the exclusive search for $p \to e^+ \pi^0$. It turns out that the best constraint arises at the moment  from the latest SK search for the two-body decay mode $p \to e^+ \pi^0$~\cite{Super-Kamiokande:2020wjk}. In fact,  this search is able to set a 90\%~CL lower limit of $185 \, {\rm GeV}$  on the effective mass scale that suppresses the relevant baryon-number violating GRSMEFT interactions. HK~measurements are expected to be able to push this limit up to $222 \, {\rm GeV}$. In~a~companion paper~\cite{inprep} we will analyse a broad range of possible laboratory probes of dimension-six and dimension-eight GRSMEFT operators, showing that the proton decay measurements discussed here set the nominally strongest bound on the effective mass scale  that suppresses the~GRSMEFT interactions.

\acknowledgments We are grateful to Maximilian Dichtl, Javi Serra and Andreas Weiler for helpful discussions and  an enjoyable collaboration leading to~\cite{inprep}. In addition, we would like to thank Roman~Zwicky for an insightful discussion on the extrapolation procedure of the LCSRs that was employed in this work. The analytical calculations in this article were performed with the help of {\tt FeynCalc}~\cite{Mertig:1990an,Shtabovenko:2016sxi,Shtabovenko:2020gxv}. Some of the Dirac traces were cross-checked against {\tt Tracer}~\cite{Jamin:1991dp}. 

\appendix

\section{Input for numerics}
\label{app:input}

In our numerical analysis we use $(m_u+m_d)/2 = (3.410 \pm 0.043) \, {\rm MeV}$~\cite{Aoki:2019cca} which corresponds to the $\overline{\rm MS}$ value of the up- and down-quark mass at $2 \, {\rm GeV}$.  Employing the two-loop renormalisation group running and the one-loop threshold corrections as implemented in {\tt RunDec}~\cite{Chetyrkin:2000yt,Herren:2017osy}, we obtain at~$1 \, {\rm GeV}$ the value $m_u+m_d = (8.60 \pm 0.11 ) \, {\rm MeV}$. By means of the Gell-Mann--Oakes--Renner~(GMOR) relation $m_\pi^2 \simeq -2 \left (m_u + m_d \right ) \braket{\bar{q}q}/f_\pi^2$~\cite{GellMann:1968rz} this value together with  the pion decay constant~$f_\pi = \left ( 130.2 \pm 0.8 \right ) \, {\rm  MeV}$~\cite{Aoki:2019cca} leads to 
\begin{equation} \label{eq:qbqcond}
\braket{\bar{q}q}  = - \big ( \left ( 256 \pm 2 \right ) \, {\rm MeV} \big ) ^3\,, 
\end{equation}
if the leading-order chiral corrections of~\cite{Bordes:2010wy} are included and uncertainties are added in quadrature. In  the case of the mixed condensate we employ~\cite{Ioffe:2002ee} 
\begin{equation} \label{eq:mixedcond}
\braket{\bar{q}g_s G \cdot \sigma q} = m_0^2 \braket{\bar{q}q} \,, \qquad m_0^2 = \left ( 0.8 \pm 0.2 \right ) \text{GeV}^2\,. 
\end{equation}
The non-perturbative parameters $\mu_\pi$ and $\rho_\pi$ can be fixed via the GMOR relation: 
\begin{equation}
\mu_\pi \equiv \frac{m_\pi^2}{m_u+m_d} \simeq - \frac{2 \braket{\bar{q}q}}{f_\pi^2} \,, \qquad \rho_\pi \equiv \frac{m_u+m_d}{m_\pi} \simeq - \frac{f_\pi^2 \hspace{0.25mm} m_\pi}{2 \braket{\bar{q}q}} \,.
\label{eq:GMOR}
\end{equation}
This leads to 
\begin{equation} \label{eq:mupirhopi}
\mu_\pi = \left (1.98 \pm 0.05 \right )  {\rm GeV} \,, \qquad 
\rho_\pi = 0.068 \pm  0.002 \,. 
\end{equation}

For the pion DAs we use the following expressions which have been derived in~\cite{Ball:1998je} (and~\cite{Braun:1989iv} in the chiral limit) with the help of a conformal expansion. One has
\begin{align}
& \phi^{(2)}(u,\mu) =  6 \hspace{0.25mm} u  \bar{u} \left[ 1+ a_2(\mu) \hspace{0.25mm} C_2^{(3/2)}(\zeta) + a_4(\mu) \hspace{0.25mm} C_4^{(3/2)} (\zeta) \right] \,, \\[3mm]
&  \phi^{(3)}_p(u,\mu) =  1 + \left( 30 \hspace{0.25mm} \eta_3(\mu) - \frac{5}{2} \hspace{0.25mm} \rho_\pi^2 \right) C_2^{(1/2)} (\zeta) \notag \\[-2mm] \\[-2mm]
& \hspace{1.95cm} + \left(-3 \hspace{0.25mm} \eta_3(\mu)\hspace{0.25mm}  \omega_3(\mu) - \frac{27}{20} \hspace{0.25mm} \rho_\pi^2 - \frac{81}{10} \hspace{0.25mm} \rho_\pi^2 \hspace{0.25mm} a_2(\mu) \right) C_4^{(1/2)}(\zeta) \,,  \notag \\[3mm]
 & \phi^{(3)}_\sigma(u,\mu) = 6 \hspace{0.25mm} u \bar{u} \left[1 + \left( 5 \hspace{0.25mm} \eta_3(\mu) - \frac{1}{2} \hspace{0.25mm} \eta_3(\mu) \hspace{0.25mm} \omega_3(\mu) - \frac{7}{20} \hspace{0.25mm} \rho_\pi^2 - \frac{3}{5}\hspace{0.25mm}  \rho_\pi^2 \hspace{0.25mm} a_2(\mu) \right) \right] C_2^{(3/2)} (\zeta) \,, \\[3mm]
  & \mathcal{T}^{(3)}\left(\alpha_d,\alpha_u,\alpha_g,\mu\right) =  360 \hspace{0.25mm} \eta_3(\mu) \hspace{0.25mm} \alpha_d \hspace{0.25mm} \alpha_u \hspace{0.25mm} \alpha_g^2 \left[1 +\frac{1}{2}\hspace{0.25mm}  \omega_3(\mu) \left(7 \alpha_g - 3\right)\right] \,, 
\end{align}
where $\bar u \equiv 1 - u$ and the expansion in terms of the Gegenbauer polynomials $C_n^{(m)} (\zeta)$ with $\zeta \equiv 2 \hspace{0.25mm} u-1$ is truncated after $n=4$. The hadronic parameters that enter the above definitions depend on the renormalisation scale $\mu$  which we set equal to~$1 \, {\rm GeV}$ in our numerical analysis.  

We adopt the numerical values of the two Gegenbauer moments given in~\cite{Khodjamirian:2011ub},
\begin{equation} \label{eq:a2a4}
a_2(1 \, {\rm GeV}) = 0.17 \pm 0.08 \,, \qquad  a_4(1 \, {\rm GeV}) = 0.06 \pm 0.10 \,,
\end{equation} 
where the moments are obtained by fitting sum rules for the electromagnetic pion form factor to the experimental data of~\cite{Huber:2008id}. For the numerical values of the other parameters we rely on the sum rules estimates presented in~\cite{Ball:2006wn}. Adding all uncertainties  in quadrature this leads to 
\begin{equation} \label{eq:f3piw3}
\eta_3(1 \, {\rm GeV}) = 0.017 \pm 0.006 \,,  \qquad \omega_3(1 \, {\rm GeV}) = -1.5 \pm 0.7 \,,
\end{equation}
when 
\beq
f_{3 \pi} (1 \, {\rm GeV}) = (0.45 \pm 0.15) \cdot 10^{-2} \, {\rm GeV}^2 \,, 
\eeq
together with the definition $f_{3 \pi} (\mu) \equiv   f_\pi \hspace{0.25mm} \mu_\pi \hspace{0.25mm}  \eta_3 (\mu)$ and~(\ref{eq:mupirhopi})  are used.

In order to obtain  the coupling strength $\lambda_p$ we use the QCD sum rule~\cite{Leinweber:1995fn}
\begin{equation}  
\lambda_p^2 = -\frac{\braket{\bar{q}q}}{16 \hspace{0.125mm} \pi^2 \hspace{0.25mm} m_p^3}  \, e^{\frac{m_p^2}{\overbar{M}^2}} \left [ \, 7 \overbar{M}^4 E_2 \left( \frac{\bar{s}_0}{\overbar{M}^2} \right) - 3 \hspace{0.25mm} m_0^2  \hspace{0.25mm} \overbar{M}^2 E_1 \left( \frac{\bar{s}_0}{\overbar{M}^2} \right)  +  \frac{19 \hspace{0.25mm} \pi^2}{18} \left \langle  \frac{\alpha_s}{\pi} \hspace{0.25mm} G^2 \right \rangle  \, \right ] \,, 
\label{eq:localsr}
\end{equation}
with $\alpha_s \equiv g_s^2/(4 \pi)$, $G^2 \equiv G_{\mu \nu}^A G^{A, \mu \nu}$ and 
\begin{equation} \label{eq:Enx}
E_n (x) \equiv 1- e^{-x} \sum_{k=0}^{n-1} \frac{x^k}{k!} \,.
\end{equation}
The parameters  $\overbar{M}$ and $\bar{s}_0$  denote the  Borel mass and the continuum threshold  of~(\ref{eq:localsr}).  The corresponding windows are $0.7 \, {\rm GeV} \leq \overbar{M} \leq 1 \, {\rm GeV}$ and $(1.4 \, \text{GeV})^2 \leq \bar{s}_0 \leq (1.5 \, \text{GeV})^2$. We furthermore employ~\cite{Ioffe:2002ee} 
\begin{equation} \label{eq:gluon2}
\left \langle  \frac{\alpha_s}{\pi} \hspace{0.25mm} G^2 \right \rangle = \left ( 0.009 \pm 0.009 \right )  \text{GeV}^4\,. 
\end{equation}
We add that using the results for the condensates and the variation of unphysical scales specified above, the sum rule~(\ref{eq:localsr}) leads to a good agreement with the LQCD results for~$\lambda_p$ (see for instance~\cite{Gavela:1988cp,Chu:1993cn,Leinweber:1994gt,RQCD:2019hps}). 

\section{Analytic results for LCSRs}
\label{app:formulaeLCSRs}

Below we  provide  the analytic expressions for the QCD correlation functions that are relevant for the calculation of the process $p \to e^+ \pi^0 G$ in the GRSMEFT. The hat on the functions $\hat{\Pi}^{\text{QCD}}_\alpha$ indicates that we have subtracted the contributions of heavy states before taking the Borel transform of the QCD results. We obtain
\begin{flalign}
\hat{\Pi}^{\text{QCD}}_{S} = \frac{ f_\pi \hspace{0.125mm}  m_p^2 }{36 \sqrt{2}} \, \Bigg\{ &\frac{\braket{\bar{q}q}}{M^2} \bigg[ \int_0^\Delta \! du \, \frac{\phi^{(2)}(u)}{\bar{u}^2} \, e(\tilde s) \left( 18 \hspace{0.125mm} \bar u - \frac{5 \hspace{0.125mm} m_0^2}{M^2}  \right)   - \frac{5 \hspace{0.125mm}  m_0^2 \hspace{0.25mm}  \phi^{(2)}(\Delta)}{Q^2 + \bar{\Delta}^2 m_\pi^2} \, e (s_0)   \bigg]  &\notag\\[-2mm] \label{eq:B1} \\[-2mm]
& \phantom{} - \frac{3 \hspace{0.125mm} \mu_\pi }{8 \pi^2} \left ( 1 - \rho_\pi^2 \right) \int_0^\Delta \! du \, \phi^{(3)}_\sigma(u) \, e(\tilde s) \hspace{0.25mm} \tilde E_1 ( \tilde{s}  ) \Bigg\} \,,& \notag
\end{flalign}

\begin{flalign}
\hat{\Pi}^{\text{QCD}}_{A_1} =  \frac{ f_\pi \hspace{0.125mm} m_p^3}{108 \sqrt{2}} \, \Bigg\{ &\frac{\braket{\bar{q}q}\hspace{-0.5mm}   \mu_\pi}{M^2}  \left ( 1 - \rho_\pi^2 \right) \bigg[ \int_0^\Delta \! du \, \frac{u \hspace{0.25mm}  \phi^{(3)}_\sigma(u)}{\bar{u}^3 M^2} \, e(\tilde s) \left(  24 \bar u - \frac{7 \hspace{0.125mm} m_0^2}{M^2} \right)    &\notag \\[2mm] 
& \hspace{-2.5cm}  + \frac{\phi^{(3)}_\sigma(\Delta)}{\bar{\Delta} \left(Q^2 + \bar{\Delta}^2 m_\pi^2 \right)} \, e (s_0) \bigg(  24  \hspace{0.125mm} \Delta \bar \Delta - \frac{7 \hspace{0.125mm} \Delta \hspace{0.125mm} m_0^2  }{M^2} - \frac{\left ( (7 + 14 \hspace{0.125mm} \Delta ) \hspace{0.125mm} \bar \Delta^3 \hspace{0.125mm}m_\pi^2 + 7 \hspace{0.125mm}\bar \Delta \hspace{0.125mm} Q^2 \right ) m_0^2 }{\ \left(Q^2 + \bar{\Delta}^2 m_\pi^2 \right)^2} \bigg) & \notag \\[2mm] 
& \hspace{-2.5cm} -  \frac{7 \hspace{0.125mm} \Delta \bar{\Delta} \hspace{0.125mm} m_0^2}{\left(Q^2 + \bar{\Delta}^2 m_\pi^2 \right)^2} \, e (s_0) \,  \frac{\partial}{\partial \Delta} \phi^{(3)}_\sigma(\Delta)    \bigg]  - \frac{9}{\pi^2}  \int_0^\Delta \! du \,  u \, \phi^{(2)} (u) \, e(\tilde s) \hspace{0.25mm} \tilde E_1 ( \tilde{s}  ) & \\[2mm]
&\hspace{-2.5cm} + \frac{72  \braket{\bar{q}q}  \mu_\pi}{M^2} \bigg[ \int_0^1 \! D\alpha\, \theta\left(\alpha-\Delta_g\right) \frac{\bar{u}\, \mathcal{T}^{(3)}\left(1-\alpha_u-\alpha_g ,\alpha_u,\alpha_g\right)}{\alpha^2 M^2}  \, e \left(\tilde{s}_g\right) &\notag \\[2mm]
& \hspace{-1.5cm} + e (s_0) \int_0^1 \! du \, d\alpha_g\, \theta \left(1-\bar{u} \alpha_g -\Delta_g \right)  \frac{\bar{u} \, \mathcal{T}^{(3)}\left(1-\bar{u}\alpha_g-\Delta_g ,\Delta_g-u\alpha_g,\alpha_g\right)}{Q^2+\Delta_g^2 \hspace{0.125mm}  m_\pi^2}\bigg] \Bigg\} \,, & \notag 
\end{flalign}

\begin{flalign}
\hat{\Pi}^{\text{QCD}}_{A_2} =  \frac{ f_\pi \hspace{0.125mm} m_p^3}{108 \sqrt{2}} \, \Bigg\{ &\frac{\braket{\bar{q}q}\hspace{-0.5mm}   \mu_\pi}{M^2}  \left ( 1 - \rho_\pi^2 \right) \bigg[ \int_0^\Delta \! du \, \frac{ \phi^{(3)}_\sigma(u)}{\bar{u}^2 M^2} \, e(\tilde s) \left(  24 \bar u - \frac{7 \hspace{0.125mm} m_0^2}{M^2} \right)    &\notag \\[2mm] 
& \hspace{-2cm}  + \frac{\phi^{(3)}_\sigma(\Delta)}{\left(Q^2 + \bar{\Delta}^2 m_\pi^2 \right)} \, e (s_0) \bigg(  24   \hspace{0.125mm} \bar \Delta - \frac{7 \hspace{0.125mm} m_0^2  }{M^2} - \frac{14  \hspace{0.125mm} \bar \Delta^3   \hspace{0.125mm} m_0^2  \hspace{0.125mm}  m_\pi^2 }{ \left(Q^2 + \bar{\Delta}^2 m_\pi^2 \right)^2} \bigg) & \notag \\[2mm] 
& \hspace{-2cm} -  \frac{7 \hspace{0.125mm} \bar{\Delta}^2  \hspace{0.125mm} m_0^2}{\left(Q^2 + \bar{\Delta}^2 m_\pi^2 \right)^2} \, e (s_0) \,  \frac{\partial}{\partial \Delta} \phi^{(3)}_\sigma(\Delta)    \bigg]  - \frac{9}{\pi^2}  \int_0^\Delta \! du \,  \bar u \, \phi^{(2)} (u) \, e(\tilde s) \hspace{0.25mm} \tilde E_1 ( \tilde{s}  ) & \\[2mm]
&\hspace{-2cm} + \frac{72\braket{\bar{q}q} \mu_\pi }{M^2} \bigg[ \int_0^1 \! D\alpha\, \theta\left(\alpha-\Delta_g\right) \frac{\bar{u}\, \mathcal{T}^{(3)}\left(1-\alpha_u-\alpha_g ,\alpha_u,\alpha_g\right)}{\alpha^2 M^2}  \, e \left(\tilde{s}_g\right) &\notag \\[2mm]
& \hspace{-1cm} + e (s_0) \int_0^1 \! du \, d\alpha_g\, \theta \left(1-\bar{u} \alpha_g -\Delta_g \right)  \frac{\bar{u} \, \mathcal{T}^{(3)}\left(1-\bar{u}\alpha_g-\Delta_g ,\Delta_g-u\alpha_g,\alpha_g\right)}{Q^2+\Delta_g^2 \hspace{0.125mm}  m_\pi^2}\bigg] \Bigg\} \,, & \notag 
\end{flalign}
\begin{flalign}
\hat{\Pi}^{\text{QCD}}_{A_3} =  \frac{ f_\pi \hspace{0.125mm} m_p}{108 \sqrt{2}} \, \Bigg\{ &\frac{\braket{\bar{q}q}\hspace{-0.5mm}   \mu_\pi}{M^2}   \bigg[  \left ( 1 - \rho_\pi^2 \right) \bigg \{  \int_0^\Delta \! du \, \frac{ \phi_\sigma^{(3)}(u)}{\bar{u}^4} \, e(\tilde s) & \notag \\[2mm] 
& \hspace{-0.5cm}\times  \bigg (  \bar u \left ( 9 -15 \hspace{0.125mm}  u + 6  \hspace{0.125mm} u^2 \right )  - \frac{6  \hspace{0.125mm} u\bar{u} \left( Q^2 + \bar{u}^2 m_\pi^2 \right) + \left ( 9 -14 \hspace{0.125mm}  u + 5  \hspace{0.125mm} u^2  \right ) m_0^2}{2M^2} &\notag \\[2mm]
& \hspace{1.2cm} + \frac{u \hspace{0.125mm} m_0^2}{M^4} \left(Q^2+\bar{u}^2 m_\pi^2 \right) \bigg)    &\notag \\[2mm] 
& \hspace{0.5cm} -\phi_\sigma^{(3)} (\Delta) \, e( s_0)  \bigg ( \frac{3 \Delta}{\bar \Delta} - \frac{ \Delta \hspace{0.25mm}  m_0^2}{\bar \Delta^2 M^2}  + \frac{7  \hspace{0.125mm} m_0^2}{2 \left(Q^2 + \bar{\Delta}^2 m_\pi^2 \right)} \bigg )    & \notag \\[1mm]
& \hspace{0.5cm} + \frac{ \Delta  \hspace{0.125mm} m_0^2}{Q^2 + \bar{\Delta}^2 m_\pi^2} \, e (s_0) \,  \frac{\partial}{\partial \Delta} \phi^{(3)}_\sigma(\Delta) \bigg \} \notag \\[2mm] 
& \hspace{0cm} -  \frac{3 \hspace{0.125mm}  m_0^2}{4} \bigg ( \int_0^\Delta \! du \, \frac{u \hspace{0.25mm} \phi^{(3)}_p(u)}{\bar{u}^2 M^2} \, e(\tilde s)  + \frac{\Delta  \hspace{0.25mm}  \phi^{(3)}_p(\Delta) }{Q^2 + \bar{\Delta}^2 m_\pi^2}  \hspace{0.25mm}  e(s_0) \bigg )   \bigg]  &\notag \\[-2mm]  \label{eq:B4} \\[-2mm]
& \hspace{0cm} + \frac{9M^2}{8\pi^2}  \int_0^\Delta \! du \, \frac{\phi^{(2)} (u)}{\bar u} \, e(\tilde s) &\notag \\[2mm]
& \hspace{0.5cm} \times \left [ \frac{u}{M^2}\left( Q^2+ \bar{u}^2 m_\pi^2 \right)  \hspace{0.25mm}  \tilde E_1 ( \tilde{s} ) - \left ( 2 - 5 u + 3 u^2 \right )  \tilde E_2 ( \tilde{s} )  \right  ]   & \notag\\[2mm]
&\hspace{0.cm} + \frac{9 \braket{\bar{q}q} \mu_\pi }{M^2} \bigg[ \int_0^1 \! D \alpha\, \theta\left(\alpha-\Delta_g\right) \frac{\mathcal{T}^{(3)}\left(1-\alpha_u-\alpha_g ,\alpha_u,\alpha_g\right)}{\alpha^3}  \, e \left(\tilde{s}_g\right) &\notag\\[2mm]
& \hspace{2.1cm} \times \left( 2 \hspace{0.125mm} \alpha \hspace{0.125mm} \bar{u} - \frac{2 \hspace{0.125mm} \bar{u} \hspace{0.125mm} Q^2}{M^2} + \frac{\alpha \hspace{0.125mm} m_\pi^2}{M^2} \left(\alpha +2\hspace{0.125mm}u-3 \right)  \right) &\notag \\[2mm]
& \hspace{0.cm} - e (s_0) \int_0^1 \! du \, d\alpha_g\, \theta \left(1-\bar{u} \alpha_g -\Delta_g \right)  \frac{\mathcal{T}^{(3)}\left(1-\bar{u}\alpha_g-\Delta_g ,\Delta_g-u\alpha_g,\alpha_g\right)}{Q^2+\Delta_g^2 \hspace{0.125mm} m_\pi^2} &\notag \\[2mm]
& \hspace{1.25cm} \times \left( \frac{2 \hspace{0.125mm} \bar{u} \hspace{0.125mm} Q^2}{\Delta_g} +m_\pi^2 \left(3-2\hspace{0.125mm} u-\Delta_g \right)  \right) \bigg] \Bigg\} \,, & \notag 
\end{flalign}
\begin{flalign}
\hat{\Pi}^{\text{QCD}}_{A_4} = -\frac{ f_\pi \hspace{0.125mm} m_p}{108 \sqrt{2}} \, \Bigg\{ &\frac{\braket{\bar{q}q}\hspace{-0.5mm}   \mu_\pi}{M^2}   \bigg[  \left ( 1 - \rho_\pi^2 \right) \bigg \{  \int_0^\Delta \! du \, \frac{ \phi_\sigma^{(3)}(u)}{\bar{u}^2} \, e(\tilde s) &\notag \\[2mm]
& \hspace{0cm} \times  \bigg ( 6 \hspace{0.125mm} \bar u + \frac{6  \left(Q^2+\bar{u}^2 m_\pi^2 \right) - 5 \hspace{0.125mm}  m_0^2}{2 \bar{u} M^2}  - \frac{m_0^2\left(Q^2+\bar{u}^2 m_\pi^2 \right)}{\bar{u} M^4}  \bigg)    &\notag \\[2mm] 
& \hspace{0cm} + \phi_\sigma^{(3)} (\Delta) \, e( s_0)  \bigg ( 3 - \frac{m_0^2}{\bar \Delta M^2}  - \frac{7  \hspace{0.125mm} m_0^2}{2 \left(Q^2 + \bar{\Delta}^2 m_\pi^2 \right)} \bigg )    & \notag \\[2mm]
& \hspace{0cm} - \frac{\bar  \Delta  \hspace{0.125mm} m_0^2}{Q^2 + \bar{\Delta}^2 m_\pi^2} \, e (s_0) \,  \frac{\partial}{\partial \Delta} \phi^{(3)}_\sigma(\Delta) \bigg \} \notag \\[2mm]
& \hspace{-0.5cm} +  \frac{3 \hspace{0.125mm}  m_0^2}{4} \bigg ( \int_0^\Delta \! du \, \frac{ \phi^{(3)}_p(u)}{\bar{u} M^2} \, e(\tilde s)  + \frac{\bar \Delta  \hspace{0.25mm}  \phi^{(3)}_p(\Delta) }{Q^2 + \bar{\Delta}^2 m_\pi^2}  \hspace{0.25mm}  e(s_0) \bigg )   \bigg]  &\notag \\[-2mm]  \label{eq:B5} \\[-2mm]
& \hspace{-0.5cm} - \frac{9}{8\pi^2}  \int_0^\Delta \! du \, \phi^{(2)} (u) \, e(\tilde s) \left [ \left( Q^2 + \bar{u}^2 m_\pi^2 \right)  \tilde E_1 ( \tilde{s} ) + 3  \hspace{0.25mm}\bar u  \hspace{0.25mm} M^2 \tilde E_2 ( \tilde{s} )  \right  ]   & \notag\\[2mm]
&\hspace{-0.5cm} + \frac{9 \braket{\bar{q}q} \mu_\pi }{M^2} \bigg[ \int_0^1 \! D\alpha\, \theta\left(\alpha-\Delta_g\right) \frac{\mathcal{T}^{(3)}\left(1-\alpha_u-\alpha_g ,\alpha_u,\alpha_g\right)}{\alpha^3}  \, e \left(\tilde{s}_g\right) &\notag  \\[2mm]
& \hspace{1.75cm} \times \left( 2 \hspace{0.125mm} \alpha \hspace{0.125mm} \bar{u} - \frac{2 \hspace{0.125mm} \bar{u}\hspace{0.125mm}  Q^2}{M^2} + \frac{\alpha^2 \hspace{0.125mm} m_\pi^2}{M^2}  \right) &\notag \\[2mm]
& \hspace{-0.5cm} - e (s_0) \int_0^1 \! du \, d\alpha_g\, \theta \left(1-\bar{u} \alpha_g -\Delta_g \right)  \frac{\mathcal{T}^{(3)}\left(1-\bar{u}\alpha_g-\Delta_g ,\Delta_g-u\alpha_g,\alpha_g\right)}{Q^2+\Delta_g^2 \hspace{0.125mm} m_\pi^2} &\notag \\[2mm]
& \hspace{0.75cm} \times \left( \frac{2 \hspace{0.125mm} \bar{u} \hspace{0.125mm} Q^2}{\Delta_g} -\Delta_g m_\pi^2 \right) \bigg] \Bigg\} \,, & \notag 
\end{flalign}
\begin{flalign}
\hat{\Pi}^{\text{QCD}}_{T_1} =  -\frac{ f_\pi}{288 \sqrt{2}} \, \Bigg\{ &\frac{\braket{\bar{q}q} \hspace{-0.25mm} m_0^2}{M^2} \bigg[ \int_0^\Delta \! du \, \frac{\phi^{(2)}(u)}{\bar{u}^3} \, e(\tilde s) \left(  \bar u - \frac{m_0^2}{M^2} \hspace{0.25mm} \bar u^2 -  \frac{ Q^2 }{M^2}  \right)    &\notag\\[-2mm] \label{eq:B6} \\[-2mm]
& \phantom{} - \frac{\phi^{(2)}(\Delta)}{\bar{\Delta}} \, e (s_0)   \bigg] + \frac{9 \hspace{0.125mm} \mu_\pi  \hspace{0.125mm}  M^2}{\pi^2}  \int_0^\Delta \! du \, \bar u \, \phi^{(3)}_p(u) \, e(\tilde s) \hspace{0.25mm} \tilde E_2 ( \tilde{s}  ) \Bigg\} \,,& \notag
\end{flalign}
\begin{flalign}
\hat{\Pi}^{\text{QCD}}_{T_2} =  \frac{ f_\pi \hspace{0.125mm}  m_p^2}{18 \sqrt{2}} \, \Bigg\{ &\frac{\braket{\bar{q}q}}{M^2} \bigg[ \int_0^\Delta \! du \, \frac{\phi^{(2)}(u)}{\bar{u}^2} \, e(\tilde s) \left(  3 \bar u - \frac{m_0^2}{M^2}  \right)   - \frac{m_0^2  \hspace{0.25mm}  \phi^{(2)}(\Delta)}{Q^2 + \bar{\Delta}^2 m_\pi^2} \, e (s_0)   \bigg]    &\notag\\[-2mm] \label{eq:B7} \\[-2mm]
& \phantom{} + \frac{3 \hspace{0.125mm} \mu_\pi }{16 \pi^2} \left ( 1 - \rho_\pi^2 \right) \int_0^\Delta \! du \, \phi^{(3)}_\sigma(u) \, e(\tilde s) \hspace{0.25mm} \tilde E_1 ( \tilde{s}  ) \Bigg\}   \,.& \notag
\end{flalign}
Including the leading pion DAs,~i.e.~the twist-2 and twist-3 contributions, we furthermore find that $\hat{\Pi}^{\text{QCD}}_{T_3} = 0$ as already discussed in Section~\ref{sec:anatomy}.

In order to write the above formulas in a compact form we have introduced
\begin{align}
& \hspace{0.5cm} \Delta  \equiv \frac{s_0+Q^2+m_\pi^2}{2 m_\pi^2} \left( 1- \sqrt{1- \frac{4 m_\pi^2 s_0}{(s_0+Q^2+m_\pi^2)^2}}\right) \,, \label{eq:def1} \\[2mm]
e(s) & \equiv   e^{-\frac{s}{M^2}} \,, \qquad \tilde E_n (s) \equiv E_n \left( \frac{s_0-s}{M^2} \right) \,, \qquad \tilde{s}  \equiv \frac{u}{\bar{u}} \hspace{-0.25mm} \left (Q^2+\bar{u} \hspace{0.15mm} m_\pi^2 \right ) \,,  \label{eq:def2}
\end{align}
\begin{align}
& \hspace{2.5cm} \Delta _g \equiv \frac{s_0+Q^2-m_\pi^2}{2 m_\pi^2} \left( \sqrt{1+\frac{4 m_\pi^2 Q^2}{(s_0+Q^2-m_\pi^2)^2}}- 1 \right) \,, \label{eq:def3} \\[2mm]
\alpha & \equiv   \alpha_u + u \alpha_g \,, \qquad D\alpha \equiv du \hspace{0.5mm} d\alpha_u  \hspace{0.5mm} d\alpha_g  \hspace{0.5mm} \theta(1-\alpha_u-\alpha_g) \,, \qquad \tilde{s}_g  \equiv \frac{\bar{\alpha}}{\alpha} \hspace{-0.25mm} \left (Q^2+\alpha \hspace{0.125mm} m_\pi^2 \right ) \,,  \label{eq:def4}
\end{align}
where the definition of $E_n (x)$ can be found in~(\ref{eq:Enx}). The upper limit of the integration over~$u$ satisfies $\Delta(Q^2=0)=1=\Delta(s_0\rightarrow \infty)$ and $\Delta \leq 1$, and it arises because the dispersion integral only has support if $s_0 \geq \tilde{s}$. We have furthermore used the shorthand notation $\bar \Delta \equiv 1- \Delta$. Similarly, one has $\Delta_g(Q^2=0)=0=\Delta(s_0\rightarrow \infty)$ and $\Delta_g \geq 0$ for the integration over the 3-particle DA.

\section{GUT-like form factors}
\label{app:check}

The hadronic form factors $w_n$  with $n=1,2,3,4$ of the process $p \rightarrow e^+ \pi^0 G$ are related to the form factors $W_{RR}^k$ with $k=0,1$ of the GUT-like decay $p \rightarrow e^+ \pi^0$. The subscript $RR$ for the latter process denotes the chiralities of the quark fields in the associated dimension-six operator
\begin{equation}
\mathcal{L}_{\slashed{B}}^{(6)} = c_{RR} \hspace{0.5mm} \epsilon^{abc} \left(d_a^T C P_R u_b \right) \left(e^T C P_R u_c\right) \,,
\label{eq:dim6}
\end{equation}
which leads to the following hadronic transition
\begin{equation}
\begin{split}
H_{RR}(p_p,q) \,u_p (p_p) &\equiv \braket{\pi^0(p_\pi) |  \, \epsilon^{abc} \left(d_a^T C P_R u_b \right) P_R u_c \,  | p(p_p)} \\[2mm]
& \equiv i P_{R} \left(W^0_{RR} (q^2) + \frac{\slashed{q}}{m_p} \, W^1_{RR}  (q^2) \right)u_p (p_p) \,,
\end{split}
\label{eq:dim6FF}
\end{equation}
where as before $q= p_p - p_\pi$. Other transitions due to operators of the type~\eqref{eq:dim6} with different chiralities for the quark fields exist and contribute to GUT-like proton decay. But only the matrix element~\eqref{eq:dim6FF}, where all quarks are right-handed, is related to the decay induced by the GRSMEFT operator~\eqref{eq:dim8}.

The hadronic matrix elements of both scenarios of proton decay, mediated by either the dimension-six term~\eqref{eq:dim6} or the dimension-eight term~\eqref{eq:dim8}, are related to the more general matrix element
\begin{equation}
H^{\alpha\beta\gamma}(p_p,q) \equiv \braket{\pi^0(p_\pi) | \epsilon^{abc} \hspace{0.125mm} d_a^\alpha  \hspace{0.125mm}  u_b^\beta  \hspace{0.125mm}  u_c^\gamma | p(p_p)} \,,
\end{equation}
where all fields are evaluated at zero. In particular, the following relations hold
\begin{align}
\left( C P_R \right)_{\alpha\beta} \, \left(P_R\right)_{\delta \gamma} \, H^{\alpha\beta\gamma}(p_p,q) &= \left[H_{RR}(p_p,q)\, u_p (p_p) \right]_\delta \,, \\[2mm]
\left( C \sigma^{\mu\nu} P_R \right)_{\alpha\beta} \, \left(P_R\right)_{\delta \gamma} \, H^{\alpha\beta\gamma}(p_p,q) &= \left[ H^{\mu\nu} (p_p,q)\, u_p (p_p) \right]_\delta \,,
\end{align}
where $H^{\mu\nu} (p_p,q)$ is defined in~\eqref{eq:onshelltensor}.

\begin{figure}[t]
\centering
\includegraphics[width=\textwidth]{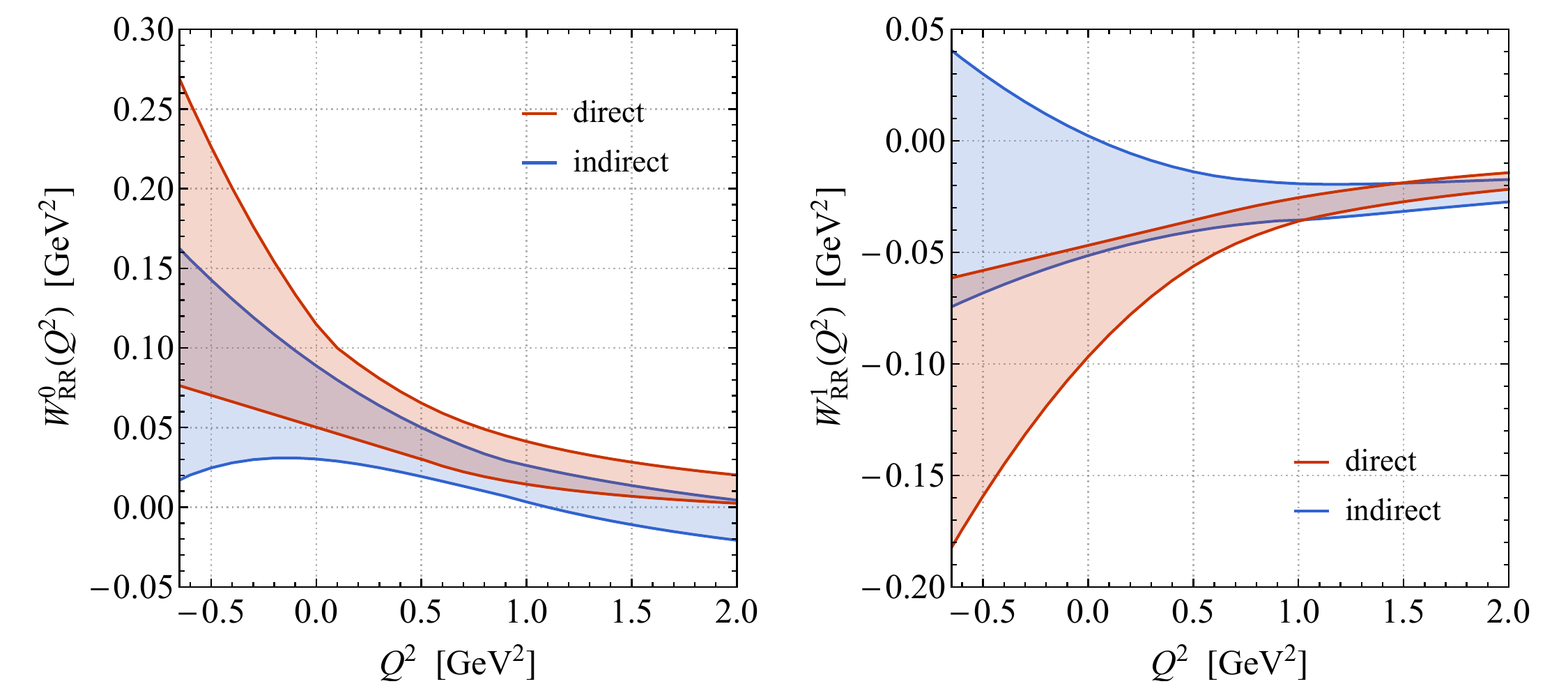}
\vspace{-6mm}
\caption{Comparison of the form factors $W_{RR}^0(Q^2)$~(left panel) and $W_{RR}^1(Q^2)$~(right panel) obtained by the direct calculation and the indirect method based on the relations~(\ref{eq:ffrel0}) and (\ref{eq:ffrel1}). The results of the direct (indirect) computation are indicated in red (blue) and the coloured envelopes include all possible solutions found in~\cite{Haisch:2021hvj} and~(\ref{eq:w1}) to~(\ref{eq:w4}), respectively. Further explanations can be found in the main text.}
\label{fig:comparison}
\end{figure}

The most general decomposition of $H^{\alpha\beta\gamma}(p_p,q)$ in terms of a set of form factors for an off-shell proton is provided in~\cite{Pire:2005ax,Pire:2021hbl} --- see in particular (4.64) of the current arXiv version of~\cite{Pire:2021hbl}. Hence, both sets of on-shell form factors $w_n$ and $W_{RR}^k$ can be related to these off-shell form factors upon using the equations of motion for the proton. In this way it is possible to relate the on-shell form factors for both scenarios of proton decay among each other. We find 
\begin{align}
W_{RR}^0 (Q^2) &= 3 \hspace{0.125mm} w_4 (Q^2) -\frac{m_p^2 - Q^2 - m_\pi^2}{ 2 m_p^2} \left( 2 \hspace{0.125mm} w_3(Q^2) - w_1(Q^2) \right) \notag \\[-2.5mm] \label{eq:ffrel0}  \\[-2.5mm]
& \phantom{xx} - \left[ \frac{Q^2}{m_p^2} + \frac{\left (m_p^2 - Q^2 - m_\pi^2 \right )^2}{4 m_p^4} \right] w_2(Q^2) \,, \notag \\[2mm] 
W_{RR}^1 (Q^2) &= 2 \hspace{0.125mm} w_3 (Q^2) - w_1(Q^2) \,.\label{eq:ffrel1}
\end{align}
Numerical results for~$W_{RR}^k(Q^2)$  were computed in our recent work~\cite{Haisch:2021hvj}, where our findings were also shown to be in agreement with the results of the state-of-the-art LQCD calculation~\cite{Aoki:2017puj} at $Q^2=0$. The form factors $w_n(Q^2)$  are derived in the same way in this work. One can therefore employ the relations~\eqref{eq:ffrel0} and~\eqref{eq:ffrel1} to directly assess the validity of first the numerical results presented in Figure~\ref{fig:ffplot} for large virtualities~($Q^2 \gg \Lambda_{\text{QCD}}^2$ with~$\Lambda_{\text{QCD}} \simeq 300 \text{ MeV}$ the QCD scale) and second the extrapolations~\eqref{eq:w1} to~\eqref{eq:w4} for physical momenta ($Q^2 \lesssim 0$). We find that the relations~\eqref{eq:ffrel0} and~\eqref{eq:ffrel1} hold numerically within uncertainties in the relevant regime $- (m_p - m_{\pi} )^2 \leq Q^2 \leq (2 \text{ GeV})^2$, even though the uncertainties of the combinations on the right-hand sides of~\eqref{eq:ffrel0} and~\eqref{eq:ffrel1} are rather large and the results tend to undershoot the more accurate results for~$W_{RR}^0 (Q^2)$ and~$W_{RR}^1 (Q^2)$ obtained by a direct calculation of the corresponding left-hand sides. This feature is illustrated by the two panels in Figure~\ref{fig:comparison}. Although the agreement is not perfect, we believe that the shown results   validate  the LCSR approach  employed in this work as well as the extrapolation procedure used to obtain~\eqref{eq:w1} to~\eqref{eq:w4}.


%

\end{document}